\documentclass[aps,groupedaddress,nofootinbib,floatfix,epfs,preprintnumbers]{revtex4}
\usepackage{amsfonts,amscd,amsmath,amssymb,graphicx,color,float,xcolor}
\usepackage[section]{placeins}
\def\ep{\text{e}}
\def\g{\mathsf{g}}
\def\oh{\frac{1}{2}}
\def\s{\mathsf{s}}

\def\k{\mathsf{k}}
\def\n{\mathsf{n}}
\def\3Q{3\text{\tiny Q}}

\def\QQbqqq{\text{\tiny Q}\bar{\text{\tiny Q}}\text{\tiny qqq}}
\def\QQqqqb{\text{\tiny QQqq}\bar{\text{\tiny q}}}

\def\QQqbqb{\text{\tiny QQ}\bar{\text{\tiny q}} \bar{\text{\tiny q}}}
\def\QbQbqq{\bar{\text{\tiny Q}}\bar{\text{\tiny Q}}\text{\tiny qq}}

\def\QQbqqb{\text{\tiny Q}\bar{\text{\tiny Q}}{\text{\tiny q}}\bar{\text{\tiny q}}}
\def\QQq{\text{\tiny QQq}}
\def\QQb{\text{\tiny Q}\bar{\text{\tiny Q}}}
\def\QQ{\text{\tiny QQ}}

\def\Qqq{\text{\tiny Qqq}}
\def\Qqb{\text{\tiny Q}\bar{\text{\tiny q}}}
\def\qQb{\text{\tiny q}\bar{\text{\tiny Q}}}
\def\Qq{\text{\tiny Qq}}

\def\qqb{\text{\tiny q}\bar{\text{\tiny q}}}
\def\nucl{\text{\tiny 3q}}
\def\qs{{\text q}}
\def\vs{{\text v}}
\def\kv{-\frac{1}{4}\ep^{\frac{1}{4}}}
\def\qsn{{\text q}_3}

\def\ws{\check{\text v}}
\def\wsz{\check{\text v}_{\text{\tiny 0}}}
\def\wso{\check{\text v}_{\text{\tiny 1}}}
\def\Vz{\bar{\text v}_{\text{\tiny 0}}}
\def\Vo{\bar{\text v}_{\text{\tiny 1}}}
\def\rq{r_q}
\def\rqb{r_{\bar q}}
\def\rv{r_v}
\def\r0{r_0}
\def\rvb{r_{\bar v}}

\def\rqqq{r_{3q}}

\begin{document}
\preprint{LMU-ASC 17/23}
\title{$Q\bar Qqqq$ Quark System, Compact Pentaquark, and Gauge/String Duality (Part II)}
\author{Oleg Andreev}
\thanks{Also on leave from L.D. Landau Institute for Theoretical Physics}
\affiliation{Arnold Sommerfeld Center for Theoretical Physics, LMU-M\"unchen, Theresienstrasse 37, 80333 M\"unchen, Germany}
\begin{abstract} 
This is the second of two companion papers in which we continue to develop the construction of the doubly heavy pentaquark systems using the gauge/string duality. In this paper, we propose a stringy description of the $Q\bar Qqqq$ system in the case of two light flavors. Our goal is to explore the lower-lying Born-Oppenheimer potentials as a function of the separation distance between the heavy quark-antiquark pair. The analysis shows that the ground state Born-Oppenheimer potential is described in terms of both hadro-quarkonia and hadronic molecules. Meanwhile a standard pentaquark configuration, which describes a genuine five-quark interaction, makes the dominant contribution to a higher lying potential. This configuration has an antiquark-diquark-diquark structure $\bar Q[qq][Qq]$ for separations larger than $0.1\,\text{fm}$. The latter enables us to establish a relation among the masses of hadrons in the heavy quark limit. To describe the structure of the potentials more clear, we define some critical separations that are related to the processes of string reconnection, breaking and junction annihilation. Additionally, we consider the generalized baryon vertices, where more than three strings can meet, and explore their implications for the pentaquark systems. 
 
 \end{abstract}
\maketitle
\vspace{0.85cm}
\section{Introduction}
\renewcommand{\theequation}{1.\arabic{equation}}
\setcounter{equation}{0}

Since the proposal of the quark model in the 1960s by Gell-Mann \cite{GM} and Zweig \cite{zweig}, the existence of exotic hadrons has remained a challenge for the physics of strong interactions. The recent observations of hidden-charm pentaquark $P_c$ states by the LHCb Collaboration \cite{Pc} has not only revived but also reinforced the longstanding interest in understanding the nature of pentaquarks \cite{ali}. These $P_c$ states are the examples of doubly heavy pentaquarks, specifically of type $Q\bar Qqqq$. In general, there may exist other pentaquark states of this type, including those with the bottom quark.

One way to handle doubly heavy quark systems is as follows. Due to the significant difference in quark masses, it appears reasonable to employ the Born-Oppenheimer (B-O) approximation, which was originally developed for use in atomic and molecular physics \cite{bo}.\footnote{For further elaboration on these ideas in the context of QCD, see \cite{braat}.} In this framework the corresponding B-O potentials are defined as the energies of stationary configurations of the gluon and light quark fields in the presence of the static heavy quark sources. The hadron spectrum is then determined by solving the Schr\"odinger equation using these potentials. 

 Lattice gauge theory is a well-established tool for studying non-perturbative QCD. Nevertheless, it still remains to be seen what it can and can't do with regard to the doubly heavy pentaquark systems. In the meantime, the gauge/string duality offers a powerful way for gaining valuable insights into this problem.\footnote{A comprehensive review of  the gauge/string duality in relation to QCD can be found in the book \cite{uaw}.} However, the existing literature notably lacks discussion on the nature of doubly-heavy pentaquarks within this framework. Bridging this gap is one of the main objectives of this paper.
 
This is the second of two companion papers in which we continue to develop the construction of the doubly heavy pentaquark systems using the gauge/string duality \cite{a-QQ3q}. The paper is organized as follows. In Sec.II, we briefly recall some preliminary results and set the framework for the convenience of the reader. Then in Sec.III, we construct and analyze a set of string configurations in five dimensions that provide a dual description of the low-lying B-O potentials in the heavy quark limit. In the process, we introduce several length scales that characterize transitions between different configurations. These length scales are in fact related to different types of string interactions, including string reconnection, breaking, and junction (baryon vertex) annihilation. In Sec.IV, we consider some aspects of gluonic excitations, with a special focus on generalized baryon vertices and their implications for the pentaquark systems. Moving on to Sec.V, we discuss a way to make the effective string model more realistic and suggest a relation among hadron masses. We conclude in Sec.VI by making a few comments on the consequences of our findings and discussing directions for future work. Appendix A contains notation and definitions. Additionally, to ensure the paper is self-contained, we include the necessary results and technical details in Appendices B and C. 
\section{Preliminaries}
\renewcommand{\theequation}{2.\arabic{equation}}
\setcounter{equation}{0}
\subsection{General procedure}

In the presence of light quarks, the B-O potentials can be determined along the lines of lattice QCD. To do this, a mixing analysis based on a correlation matrix is necessary, as explained in \cite{FK} in the case of string breaking. The diagonal elements of this matrix correspond to the energies of stationary string configurations, while the off-diagonal elements describe transitions between these configurations. The potentials can then be determined by calculating the eigenvalues of the matrix.

\begin{figure}[htbp]
\centering
\includegraphics[width=9cm]{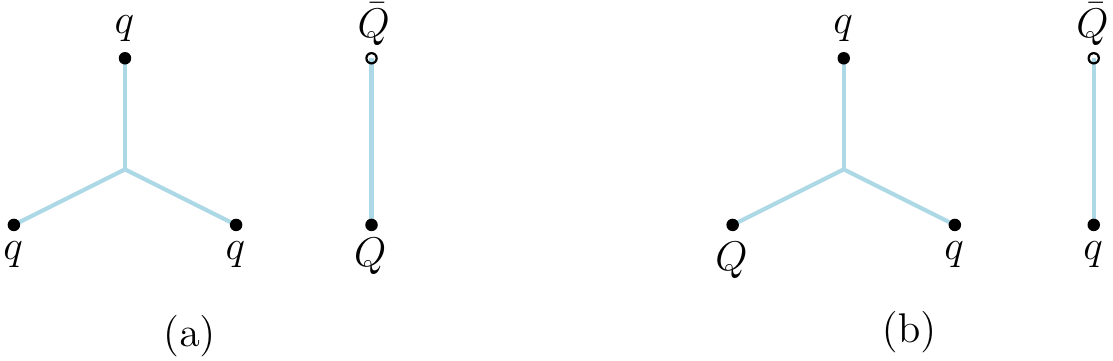}
\caption{{\small Basic string configurations. Here and later, non-excited strings are denoted by straight lines.}}
\label{c40}
\end{figure}

Now consider the $Q\bar Qqqq$ quark system and examine the corresponding string configurations within the four-dimensional string models \cite{XA}. In our discussion, we'll assume $N_f=2$, which means there are two dynamical flavors with equal mass ($u$ and $d$ quarks).\footnote{Extending the analysis to $N_f=2+1$, by including the $s$ quark, is straightforward.} First, let's examine string configurations with only the valence quarks. These are the basic configurations shown in Figure \ref{c40}. Each configuration consists of the valence quarks and antiquark 

\noindent connected by the strings. Three strings may join at a point known as the string junction \cite{XA2}.\footnote{Notably, it took almost a quarter of a century to discover the initial evidence of the string junction in numerical simulations. See, for example, \cite{lattice345}.} These configurations are disconnected and look like non-interacting hadrons. Clearly, the latter is true only if the hadrons are well-separated. If they are not, configuration (a) describes a hadro-quarkonium state, a $Q\bar Q$ pair in a nucleon cloud, and configuration (b) describes a hadron molecule.  

\begin{figure}[H]
\centering
\includegraphics[width=3cm]{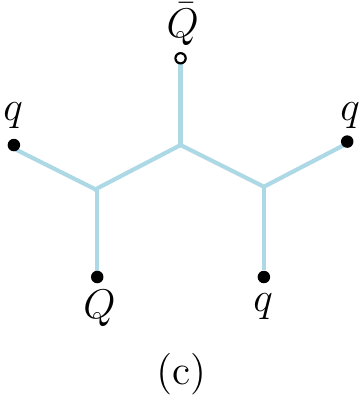}
\caption{{\small A pentaquark configuration.}}
\label{c41}
\end{figure}

To get further, we assume that other (excited) configurations can be constructed by adding extra string junctions and virtual quark-antiquark pairs to the basic configurations. This also results in an increased number of strings and, therefore, intuitively indicates that these configurations possess higher energies. So to some extent, the junctions and $q\bar q$ pairs can be thought of as kinds of elementary excitations. For our purposes, relatively simple configurations suffice. In particular, adding a pair of junctions to the basic configurations results in the pentaquark configuration illustrated in Figure \ref{c41}. Since it describes the genuine five-body interaction of quarks, we call it the pentaquark configuration.\footnote{As we will see in Sec.III, such a configuration makes the dominant contribution to one of the B-O low-lying potential at small heavy quark separations. Because of this, we will add the word "compact" as a prefix.} 

\begin{figure}[htbp]
\centering
\includegraphics[width=16cm]{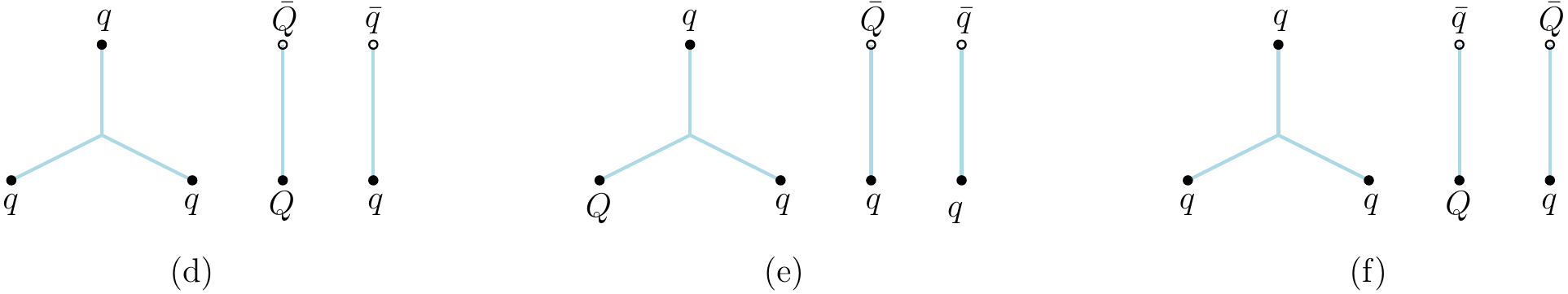}
\caption{{\small String configurations with one virtual quark pair.}}
\label{c42}
\end{figure}

Similarly, adding one $q\bar q$ pair results in the configurations shown in Figure \ref{c42}. The configurations (d) and (e) are simple modifications of the configurations (a) and (b), respectively. The configuration (f) is obtained from those by quark exchange. One can interpret configuration (d) as a hadro-quarkonium state, namely a $Q\bar Q$ pair surrounded by a pion-nucleon cloud, while the other configurations can be interpreted as hadron molecules within pion and nucleon clouds. It is noteworthy that other elementary excitations may be involved. We return to this issue in Sec.IV.

\begin{figure}[H]
\centering
\includegraphics[width=15.75cm]{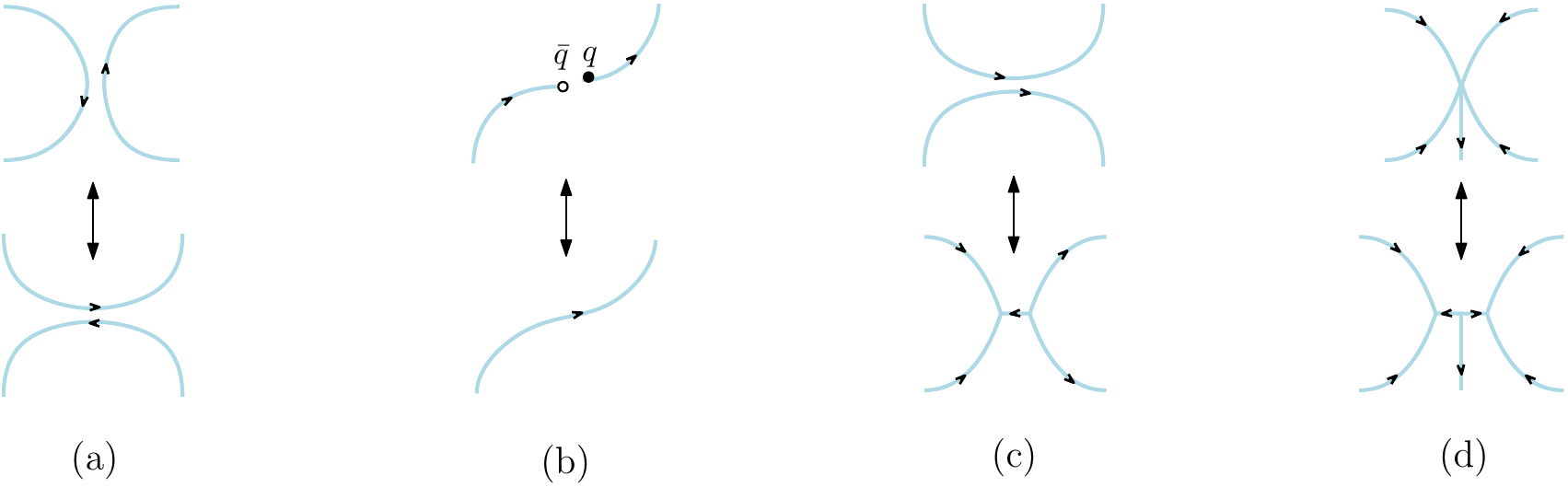}
\caption{{\small Some string interactions: (a) reconnection, (b) breaking, (c) junction annihilation, (d) junction fusion.}}
\label{sint}
\end{figure}

The transitions between the configurations arise due to string interactions. In Figure \ref{sint}, we sketch four different

\noindent  types of interactions which will be discussed in the following sections. This is part of the big picture of QCD strings. Later on, we will introduce the notion of a critical separation between the heavy quarks, which characterizes each interaction. This is helpful for gaining a deeper understanding of the physics of QCD strings and the structure of B-O potentials.

\subsection{A short account of the five-dimensional string model}

In our study of the $Q\bar Qqqq$ system, we will use the formalism recently developed in \cite{a-strb}. This formalism is general and can be adapted to any model of AdS/QCD, although we illustrate it by performing calculations in one of the simplest models. 

For the purposes of this paper, we consider a five-dimensional Euclidean space with a metric

\begin{equation}\label{metric}
ds^2=\ep^{\s r^2}\frac{R^2}{r^2}\Bigl(dt^2+(dx^i)^2+dr^2\Bigr)
\,,
\end{equation}
where $r$ is the fifth dimension of the space. Such a space represents a deformation of the Euclidean $\text{AdS}_5$ space of radius $R$, with a deformation parameter $\s$. The boundary is at $r=0$, and the so-called soft wall at $r=1/\sqrt{\s}$. This model is particularly appealing due to its relative computational simplicity and its potential for phenomenological applications. Here let us just mention that the model of \cite{az1} provides a good fit to the lattice data obtained for the heavy quark potential \cite{white}.\footnote{See also \cite{a-3qPRD} for another good example.}
 
To construct the string configurations of Figures \ref{c40}-\ref{c42} in five dimensions, we need certain building blocks. The first is a Nambu-Goto string governed by the action 

\begin{equation}\label{NG}
S_{\text{\tiny NG}}=\frac{1}{2\pi\alpha'}\int d^2\xi\,\sqrt{\gamma^{(2)}}
\,.
\end{equation}
Here $\gamma$ is an induced metric, $\alpha'$ is a string parameter, and $\xi^i$ are world-sheet coordinates. 

The second is a high-dimensional counterpart of the string junction, known as the baryon vertex. In the AdS/CFT correspondence, this vertex is 
    supposed to be a dynamic object which is a five brane wrapped on an internal space $\mathbf{X}$ \cite{witten}, and correspondingly the antibaryon vertex is an antibrane. Both objects look point-like in five dimensions. In \cite{a-3qPRD} it was observed that the action for the baryon vertex, written in the static gauge, 

\begin{equation}\label{baryon-v}
S_{\text{vert}}=\tau_v\int dt \,\frac{\ep^{-2\s r^2}}{r}
\,
\end{equation}
yields very satisfactory results, when compared to the lattice calculations of the three-quark potential. Note that $S_{\text{vert}}$ represents the worldvolume of the brane if $\tau_v={\cal T}_5R\,\text{vol}(\mathbf{X})$, with ${\cal T}_5$ the brane tension. Unlike AdS/CFT, we treat $\tau_v$ as a free parameter to account for $\alpha'$-corrections as well as the possible impact of other background fields.\footnote{Similar to AdS/CFT, there is an expectation of the presence of an analogue of the Ramond-Ramond fields on $\mathbf{X}$.} In the case of zero baryon chemical potential, it is natural to suggest the same action for the antibaryon vertex, such that $S_{\bar{\text{vert}}}=S_{\text{vert}}$.

To model the two light quarks of equal mass, we introduce a background scalar field $\text{T}(r)$, as proposed in \cite{son}. This scalar field couples to the worldsheet boundary as an open string tachyon $S_{\text{q}}=\int d\tau e\text{T}$, where $\tau$ is a coordinate on the boundary and $e$ is a boundary metric (an einbein field). Thus, the light quarks are at string endpoints in the interior of five-dimensional space. For our purposes, we only consider a constant field $\text{T}_0$ and worldsheets with straight-line boundaries in the $t$-direction. In this case, the action written in the static gauge can be expressed as

\begin{equation}\label{Sq}
S_{\text q}=\text{T}_0R\int dt \frac{\ep^{\oh\s r^2}}{r}
\,
\end{equation}
and recognized as the action of a point particle of mass ${\text T}_0$ at rest.\footnote{The masses of the light quarks can be determined by fitting the string breaking distance for the $Q\bar Q$ system to the lattice data of \cite{bulava}, which yields $m_{u/d}=46.6,\text{MeV}$ \cite{a-stb3q} for the parameter values used in this paper.} Clearly, at zero baryon chemical potential the same action also describes the light antiquarks, and thus $S_{\bar{\text q}}=S_{\text q}$. 

It's worth noting the visual analogy between tree level Feynman diagrams and static string configurations. In the language of Feynman diagrams, the building blocks mentioned above respectively play the roles of propagators, vertices, and tadpoles.

\section{The string theory analysis in five dimensions}
\renewcommand{\theequation}{3.\arabic{equation}}
\setcounter{equation}{0}

Now we will describe the $Q\bar Qqqq$ system in five dimensions. Our basic approach is as follows: following the hadro-quarkonium picture \cite{voloshin}, we consider the light quarks as clouds, and therefore, it only makes sense to speak about their average positions, or equivalently, the centers of the clouds. The heavy quarks are point-like objects inside the clouds. Our goal is to determine the low-lying B-O potentials as a function of the distance between the heavy quark and antiquark.

We begin our discussion with the basic string configurations and then move on to the remaining ones before ending with the potentials. To get an intuitive idea of what a configuration looks like in five dimensions, one can place it on the boundary of five-dimensional space. A gravitational force pulls the light quarks and strings into the interior, while the heavy (static) quarks remain at rest. This mostly helps, but there are some exceptions. We will see shortly that the shape of several configurations changes with the separation between the heavy quarks, making the problem more complicated.

\subsection{The disconnected configurations (a) and (b)}

\begin{figure}[H]
\centering
\includegraphics[width=5.25cm]{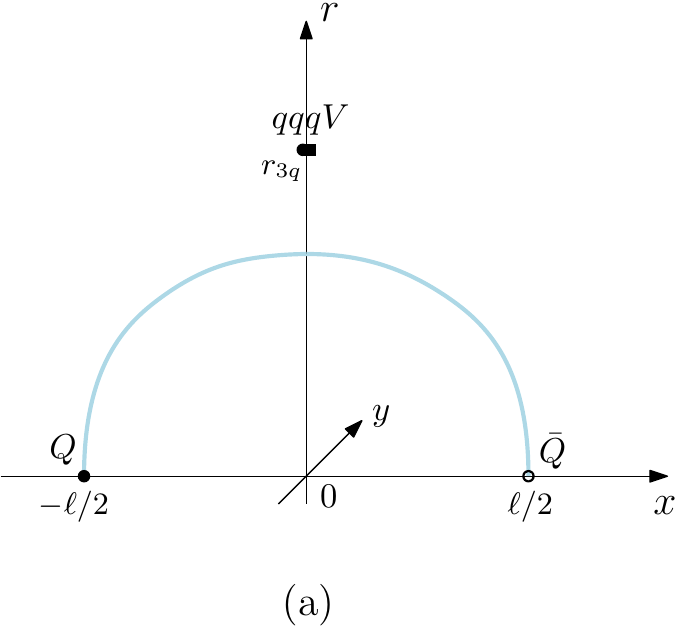}
\hspace{2.5cm}
\includegraphics[width=5.25cm]{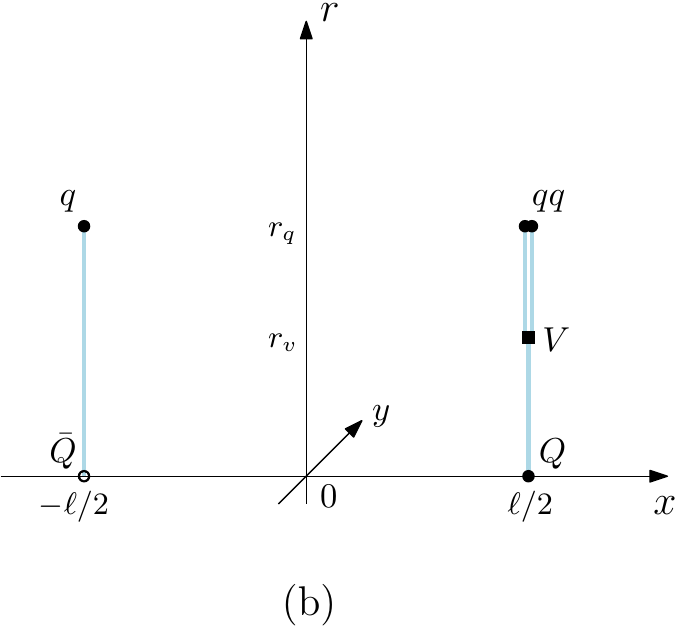}
\caption{{\small The basic configurations in five dimensions. The heavy quark and antiquark are placed on the boundary at $r=0$ and are separated by a distance of $\ell$.
The light quarks, baryon vertices and nucleon are in the interior at $r=\rq$, $r=\rv$, and $r=r_{3q}$, respectively.}}
\label{con-ab}
\end{figure}

Consider configuration (a), which can be interpreted as a $Q\bar Q$ pair in a nucleon cloud. In the following discussion, we will average over all possible nucleon positions. It was observed in \cite{pion-factor} that in the case of a pion cloud, the total energy is almost equal to the sum of the rest energies of two hadrons. We will assume that such a factorization also holds in the present case.\footnote{In general, the factorization takes place if the hadrons are far apart, but for shorter distances they do interact with each other. The partial answer to this will become clear from the construction of the B-O potentials via a matrix Hamiltonian, where hadronic interactions are encoded in the off-diagonal terms of the Hamiltonian.} In five dimensions, the configuration looks like the one shown in Figure \ref{con-ab}(a), consisting of   
\noindent two parts: the lower corresponds to the $Q\bar Q$ system, and the upper to the nucleon. The total energy is the sum of two terms 

\begin{equation}\label{Ea}
E^{\text{(a)}}=E_{\QQb}+E_{\nucl}
\,.	
\end{equation}
In the static limit, $E_{\QQb}$ was computed in \cite{az1}.\footnote{For convenience, we give a brief summary of the results in Appendix B.} Meanwhile, $E_{\nucl}$ was computed in \cite{a-QQ3q} and is given by 

\begin{equation}\label{nucl}
	E_{\nucl}=
	3\g\sqrt{\frac{\s}{\qsn}}
	\Bigl(
\k\ep^{-2\qsn}
+\n\ep^{\oh\qsn}
\Bigr)
	\,.
\end{equation}
Here $\g=\frac{R^2}{2\pi\alpha'}$, $\k=\frac{\tau_v}{3\g}$, $\n=\frac{\text{T}_0 R}{\g}$, and $\qsn$ is a solution to the equation 

\begin{equation}\label{3q}
\k(1+4q_3)+\n(1-q_3)\ep^{\frac{5}{2}q_3}=0
\,.
\end{equation}	
This equation is the force balance equation in the $r$-direction. It is derived by varying the action $S=S_{\text{v}}+3S_{\text{q}}$ with respect to $\rqqq$. Note that $q_3=\s\rqqq^2$.

Now let's consider configuration (b). Again, the total energy is just the sum of the rest energies 

\begin{equation}\label{Eb}
E^{\text{(b)}}=E_{\qQb}+E_{\Qqq}
\,.	
\end{equation}
The first term is the rest energy of a heavy-light meson which equals to $E_{\Qqb}$ at zero baryon chemical potential. The latter computed in \cite{a-strb} is

\begin{equation}\label{Qqb}
E_{\Qqb}=\g\sqrt{\s}\Bigl({\cal Q}(\qs)+\n \frac{\ep^{\oh \qs}}{\sqrt{\qs}}\Bigr)+c
\,,
	\end{equation}
where the function ${\cal Q}$ is defined in Appendix A, $c$ is a normalization constant, and  $\qs$ is a solution to the equation 

\begin{equation}\label{q}
\n(q-1)+\ep^{\frac{q}{2}}=0
\,
\end{equation}
in the interval $[0,1]$. This equation is nothing else but the force balance equation in the $r$-direction and is derived by varying the action $S=S_{\text{\tiny NG}}+S_{\text q}$ with respect to $\rq$. Note that $q=\s \rq^2$. 

The second term represents the rest energy of a heavy-light baryon. It was also computed in \cite{a-strb}, with the result 

\begin{equation}\label{EQqq}
E_{\Qqq}=\g\sqrt{\s}\Bigl(2{\cal Q}(\qs)-{\cal Q}(\vs)
+2\n \frac{\ep^{\oh \qs}}{\sqrt{\qs}}
+3\k \frac{\ep^{-2\vs}}{\sqrt{\vs}}
\Bigr)+c
\,.
	\end{equation}
Here $\vs$ a solution to the equation

\begin{equation}\label{v}
1+3\k(1+4v)\ep^{-3v}=0
\,
\end{equation}
and $v=\s\rv^2$. The above equation is the force balance equation in the $r$-direction at $r=\rv$. It is derived by varying the action $S=3S_{\text{\tiny NG}}+2S_{\text q}+S_{\text{vert}}$ with respect to $\rv$. 

We conclude our discussion of the basic configurations with some remarks. Firstly, it was shown in \cite{a-QQq} that in the interval $[0,1]$, Eq.\eqref{v} has solutions if and only if $\k$ is restricted to the range $-\frac{\ep^3}{15} <\k\leq -\frac{1}{4}\ep^{\frac{1}{4}}$. In particular, there exists a single solution $\vs=\frac{1}{12}$ at $\k= -\frac{1}{4}\ep^{\frac{1}{4}}$. Secondly, the analysis of configuration (b) assumes that $\vs\leq \qs$. Although this is not true for all possible parameter values, it definitely is for those we use to make predictions. Finally, the solutions $\qs$ and $\vs$ are associated with the light quarks and baryon vertices, and as such, they are independent of the separation of the heavy quarks.
\subsection{The connected configuration (c)}

Having understood the basic string configurations, we can now discuss the pentaquark configuration (c). In doing so, it is natural to suggest that if a configuration contributes to the ground state, or at least to one of the low excited states, its shape is dictated by symmetry. For the configuration at hand, the most symmetric case involves placing all the light quarks in the middle between the heavy quark sources. This is a good starting point for small separations. At larger separations, the pentaquark configuration  does change shape, as we will see shortly.
 
\subsubsection{Small $\ell$}

\begin{figure}[H]
\centering
\includegraphics[width=6.6cm]{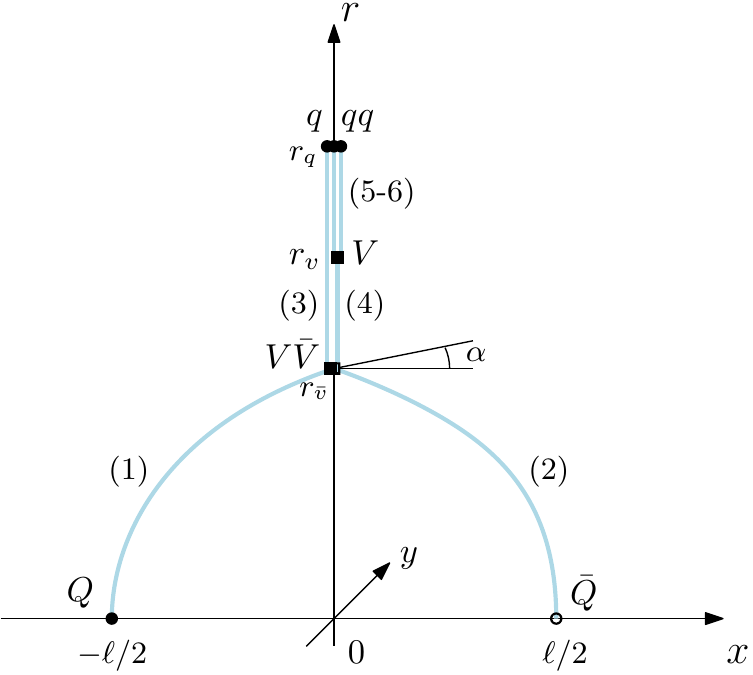}
\caption{{\small The pentaquark configuration for small $\ell$. The light quarks and baryon vertices are on the $r$-axis at $r=\rq$, $r=\rv$, and $r=\rvb$. Here and later, $\alpha$ represents the tangent angle at the endpoint of the first string.}}
\label{c51}
\end{figure}

In this case the corresponding string configuration is depicted in Figure \ref{c51}.  From a four-dimensional perspective
\noindent the light quarks are located in the middle between the heavy ones. It is assumed that $\rq$, $\rv$, and $\rvb$ satisfy the condition $\rq>\rv>\rvb$, which is indeed true for the parameter values we are using. 

The total action is the sum of the Nambu-Goto actions plus the actions for the vertices and light quarks

\begin{equation}\label{Sc1}
S=\sum_{i=1}^6 S_{\text{\tiny NG}}^{(i)}+3S_{\text{vert}}+3S_{\text q}
\,.
\end{equation}

If one chooses the static gauge $\xi^1=t$ and $\xi^2=r$ for the Nambu-Goto actions and considers the $x$'s as a function of $r$, then the boundary conditions for them are 

\begin{equation}\label{boundary-s}
x^{(1;2)}(0)=\mp\oh\ell\,,
\qquad
x^{(1,2,3,4)}(\rvb)=x^{(4,5,6)}(\rv)=x^{(3,5,6)}(\rq)=0\,.
\end{equation}
Now the action takes the form\footnote{We drop the subscript $(i)$ when it does not cause confusion.}

\begin{equation}\label{Sc11}
S=\g T
\biggl(
2\int_{0}^{\rvb} \frac{dr}{r^2}\,\ep^{\s r^2}\sqrt{1+(\partial_r x)^2}\,\,
+
\int_{\rvb}^{\rq} \frac{dr}{r^2}\,\ep^{\s r^2}
+
\int_{\rvb}^{\rv} \frac{dr}{r^2}\,\ep^{\s r^2}
+
2\int_{\rv}^{\rq} \frac{dr}{r^2}\,\ep^{\s r^2}
+
6\k\,\frac{\ep^{-2\s\rvb^2}}{\rvb}
+
3\k\,\frac{\ep^{-2\s\rv^2}}{\rv}
+
3\n\frac{\ep^{\frac{1}{2}\s\rq^2}}{\rq}
\,\biggr)
\,,
\end{equation}
where $T=\int dt$ and $\partial_rx=\frac{\partial x}{\partial r}$. We set $x=const$ for all the strings stretched along the $r$-axis. The integrals represent the contributions of the strings, while the remaining terms represent the contributions of the vertices and light quarks. 

To find a stable configuration, we extremize the action with respect to $x$, which describes the profiles of strings (1) and (2), and with respect to $\rvb$, $\rv$, and $\rq$, which describe the locations of the vertices and light quarks. As explained in Appendix B of \cite{a-stb3q}, varying with respect to $x$ gives the expressions for the separation distance and the energy of the strings

\begin{equation}\label{lc-s}
\ell=\frac{2}{\sqrt{\s}}{\cal L}^+(\alpha,\bar v)
\,,
\qquad
E^{(1,2)}=\g\sqrt{\s}\,{\cal E}^{+}(\alpha, \bar v)+c
\,.
\end{equation}
Here $c$ is the normalization constant as before. It is easy to see that varying the action with respect to $\rq$ and $\rv$ leads to Eqs.\eqref{q} and \eqref{v}. Putting all together, we find 
 
 \begin{equation}\label{Ec-s}
E^{\text{(c)}}=\frac{S}{T}=E_{\QQbqqq}=\g\sqrt{\s}
\biggl(
2{\cal E}^+(\alpha,\bar v)
+
3{\cal Q}(\qs)-2{\cal Q}(\bar v)-{\cal Q}(\vs)
+
6\k\frac{\ep^{-2\bar v}}{\sqrt{\bar v}}
+
3\k\frac{\ep^{-2\vs}}{\sqrt{\vs}}
+
3\n\frac{\ep^{\oh\qs}}{\sqrt{\qs}}
\,
\biggr)
+2c
\,.
\end{equation}
We have used the fact that $\int_a^b\frac{dx}{x^2}\ep^{cx^2}=\sqrt{c}\bigl({\cal Q}(cb^2)-{\cal Q}(ca^2)\bigr)$. Here, $\bar v=\s\rvb^2$, and the functions ${\cal L}^+$ and ${\cal E}^+$ are defined in Appendix A. Finally, varying with respect to $\rvb$ leads to the equation

\begin{equation}\label{alpha1}
\sin\alpha=1+3\k(1+4\bar v)\ep^{-3\bar v}
\,,	
\end{equation}
which is nothing else but the force balance equations in the $r$-direction at $r=r_{\bar v}$.

Thus, the energy of the pentaquark configuration is given parametrically by $E_{\QQbqqq}=E_{\QQbqqq}(\bar v)$ and $\ell=\ell(\bar v)$, where the parameter $\bar v$ varies from 0 to $\vs$. The lower limit is determined by $\ell(0)=0$, and the upper limit by $\bar v=\vs$, which corresponds to the situation where string (4) shrinks into a point. 
\subsubsection{Slightly larger $\ell$}

\begin{figure}[H]
\centering
\includegraphics[width=6.75cm]{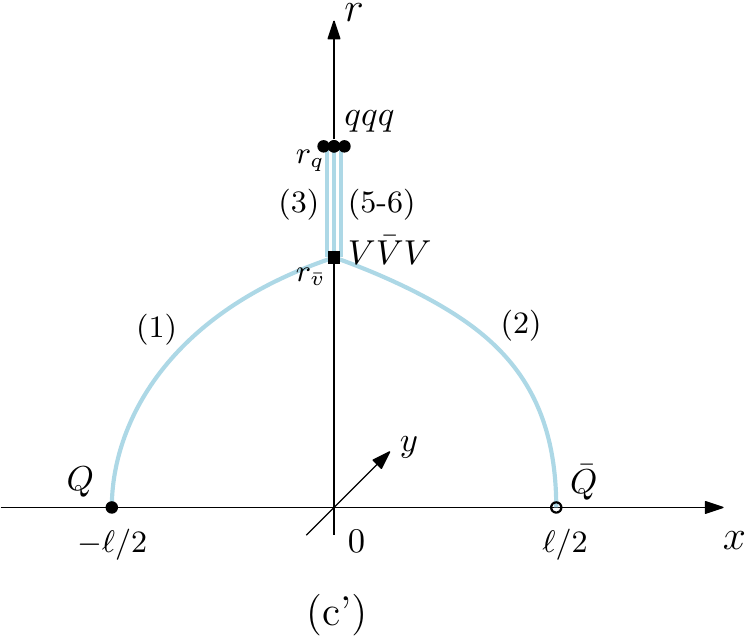}
\hspace{2.5cm}
\includegraphics[width=6.75cm]{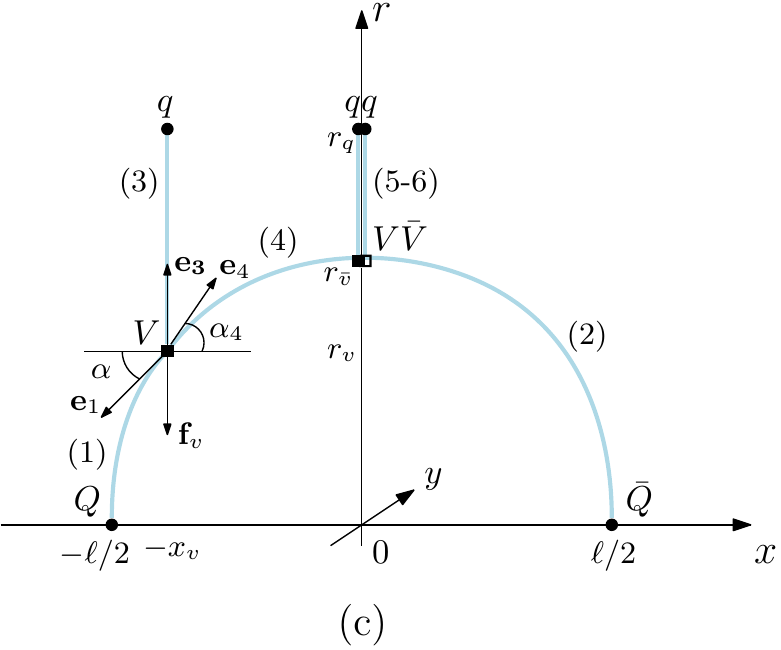}
\caption{{\small Left: The configuration of Figure \ref{c51} at $\bar v=\vs$. Right: The pentaquark configuration for $\ell$ ranging from $\ell(\vs)$ to $\ell(\qs)$.}}
\label{c52}
\end{figure}

A straightforward numerical analysis of \eqref{lc-s} shows that $\ell(\bar v)$ increases monotonically and remains finite at $\bar v = \vs$. This implies that the $V\bar V$ pair gradually moves deeper into the bulk until it reaches the baryon vertex $V$, whose position is independent of the separation between the heavy quarks. As a result, the configuration becomes that of Figure \ref{c52}(c'), where string (4) has collapsed to a point. It turns out that proceeding further with such a configuration 
\noindent is impossible. As explained in Appendix C, it only exists for separations slightly exceeding $\ell(\vs)$. A possible way out is to consider another configuration in which the vertices are spatially separated as depicted in Figure \ref{c52}(c). It can be obtained from configuration (c') by splitting the baryon vertices and stretching a string between them. 

Formally, this configuration is also governed by the action \eqref{Sc1}, but with the boundary conditions replaced by 

\begin{equation}\label{boundary-s2}
x^{(1;2)}(0)=\mp\oh\ell\,,
\qquad
x^{(1,3,4)}(\rv)=x^{(3)}(\rq)=-x_v\,,
\qquad
x^{(2,4,5,6)}(\rvb)=x^{(5,6)}(\rq)=0\,.
\end{equation}
So it now reads 

\begin{equation}\label{Sc2}
\begin{split}
S=\g T
\biggl(
&\int_{0}^{\rv} \frac{dr}{r^2}\,\ep^{\s r^2}\sqrt{1+(\partial_r x)^2}
+
\int_{0}^{\rvb} \frac{dr}{r^2}\,\ep^{\s r^2}\sqrt{1+(\partial_r x)^2}
+
\int_{\rv}^{\rq} \frac{dr}{r^2}\,\ep^{\s r^2}
+
\int_{\rv}^{\rvb} \frac{dr}{r^2}\,\ep^{\s r^2}\sqrt{1+(\partial_r x)^2}\\
+2 &\int_{\rvb}^{\rq} \frac{dr}{r^2}\,\ep^{\s r^2} 
+
3\k\,\frac{\ep^{-2\s\rv^2}}{\rv}
+
6\k\,\frac{\ep^{-2\s\rvb^2}}{\rvb}
+
3\n\frac{\ep^{\frac{1}{2}\s\rq^2}}{\rq}
\,\biggr)
\,.
\end{split}
\end{equation}
Here we set $x^{(3,5,6)}=const$. The integrals correspond to the contributions of strings (1)-(6), respectively.

Given the action, it is straightforward to extremize it with respect to $x_v$ and $\rv$, which describe the location of the single baryon vertex. The result can be conveniently expressed in a vector form as follows

\begin{equation}\label{V-fbe}
\mathbf{e}_1+\mathbf{e}_3+\mathbf{e}_4+\mathbf{f}_v=0
\,,
\end{equation}
where $\mathbf{e}_1=\g w(\rv)(-\cos\alpha,-\sin\alpha)$, $\mathbf{e}_3=\g w(\rv)(0,1)$, $\mathbf{e}_4=\g w(\rv)(\cos\alpha_4,\sin\alpha_4)$, and $\mathbf{f}_v=(0,-3\g\k\,\partial_{\rv}\frac{\ep^{-2\s\rv^2}}{\rv})$, with  $w(r)=\ep^{\s r^2}/r^2$ and $\alpha_i\leq\frac{\pi}{2}$. This is the force balance equation at the vertex position, as shown in Figure \ref{c52}(c). Its $x$-component reduces to  

\begin{equation}\label{Vx-fbe}
\cos\alpha-\cos\alpha_4 =0
\,. 
\end{equation}
Since the equation has a straightforward solution $\alpha_4=\alpha$, it implies that strings (1) and (4) are smoothly joined together to form a single string, which we refer to as string (1). The vertex, therefore, does not affect the string.\footnote{In fact, this is true only for $r_v=\sqrt{\vs/\s}\,$ as follows from the $r$-component of the force balance equation.} If so, then the $r$-component becomes equivalent to Eq.\eqref{v} whose solution is given by $\vs$. As a result, the action takes the form  

\begin{equation}\label{actionIIc3}
S=\g T
\biggl(
2 \int_{0}^{\rvb} \frac{dr}{r^2}\,\ep^{\s r^2}\sqrt{1+(\partial_r x)^2}
+
\int_{r_{\vs}}^{\rq} \frac{dr}{r^2}\,\ep^{\s r^2}
+2\int_{\rvb}^{\rq} \frac{dr}{r^2}\,\ep^{\s r^2} 
+
3\k\,\frac{\ep^{-2\s r_{\vs}^2}}{r_{\vs}}
+
6\k\,\frac{\ep^{-2\s\rvb^2}}{\rvb}
+
3\n\frac{\ep^{\frac{1}{2}\s\rq^2}}{\rq}
\,\biggr)
\,.
\end{equation}
Here the first integral corresponds to the contributions of string (1)-(2), and $r_{\vs}=\sqrt{\vs/\s}$. Note that varying the action with respect to $\rq$ and $\rvb$ results respectively in Eqs.\eqref{q} and \eqref{alpha1}. 

By essentially the same arguments that we gave for the expression \eqref{Ec-s}, the energy of this configuration can be written as 

\begin{equation}\label{Ec-s2}
E_{\QQbqqq}
=
2\g\sqrt{\s}
\biggl(
{\cal E}^+(\alpha,\bar v)
+
{\cal Q}(\qs)
-
{\cal Q}(\bar v)
+
3\k\frac{\ep^{-2\bar v}}{\sqrt{\bar v}}
+
\n\frac{\ep^{\oh\qs}}{\sqrt{\qs}}
\biggr)
+
E_0+2c
\,,
\end{equation}
where $E_0=\g\sqrt{\s}\bigl({\cal Q}(\qs)-{\cal Q}(\vs)+3\k\frac{\ep^{-2\vs}}{\sqrt{\vs}}+\n\frac{\ep^{\oh\qs}}{\sqrt{\qs}}\bigr)$.  The parameter $\bar v$ takes values in the interval $[\vs,\qs]$.

At this point, two remarks are in order. Firstly, as seen from Figure \ref{c52}(c), the spatial positions of the light quarks along the $x$-axis suggest an antiquark-diquark-diquark $\bar Q[qq][Qq]$ structure.\footnote{In fact, the separation between the $Q$ and $q$ (attached to string (3)) quarks decreases as the heavy quark separation increases.} Such a structure was assumed in \cite{maiani} and was found to be phenomenologically useful. Secondly, it was demonstrated in \cite{a-QQqbqb} that the connected tetraquark configuration for the $\bar Q\bar Q qq$ system has an antiquark-antiquark-diquark $\bar Q\bar Q [qq]$ structure. Since the diquark $[Qq]$ is color-antitriplet, it is reasonable to assume that there exists a relation between the energies of the pentaquark and tetraquark configurations. A closer inspection shows that this is indeed the case. The first term in \eqref{Ec-s2} is equal to the energy of the tetraquark configuration \cite{a-QQqbqb}, and thus the energies are just shifted by a constant equal to the second term. Explicitly, 

\begin{equation}\label{maiani}
	E_{\QQbqqq}(\ell)=E_{\QQqbqb}(\ell)+E_0 
	\qquad \text{for}\quad\ell\geq \ell(\vs)\,.
\end{equation}
We have used the fact that $E_{\QbQbqq}=E_{\QQqbqb}$ at zero baryon chemical potential.

\subsubsection{Intermediate and large $\ell$}

\begin{figure}[H]
\centering
\includegraphics[width=6.75cm]{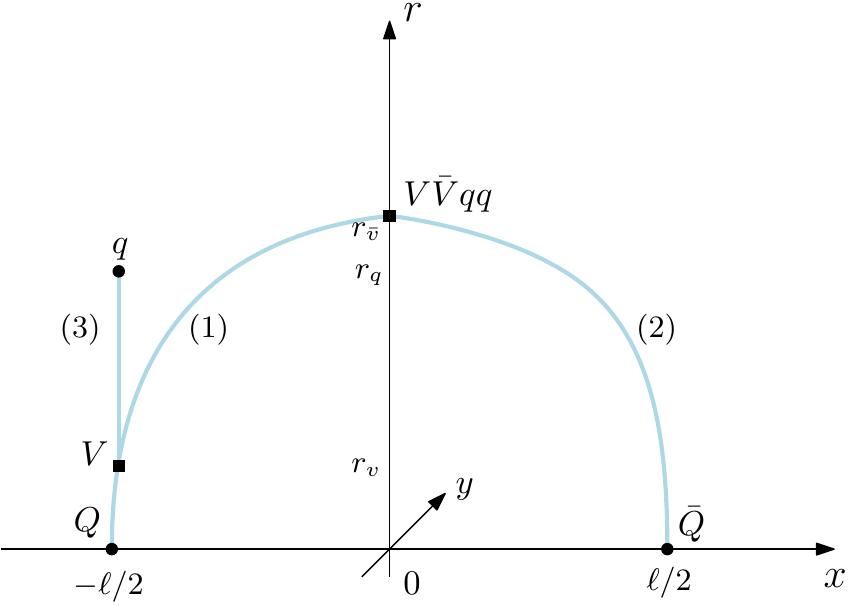}
\hspace{2.5cm}
\includegraphics[width=6.75cm]{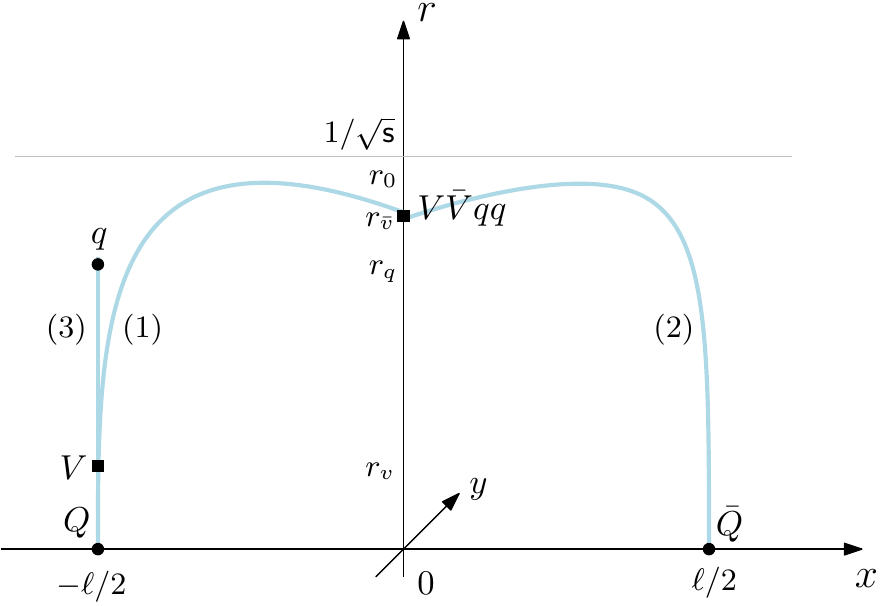}
\caption{{\small The pentaquark configuration for intermediate (left) and large (right) heavy quark separations. The horizontal line represents the soft wall at $r=1/\sqrt{\s}$.}}
\label{c53}
\end{figure}

Numerical analysis shows that $\ell(v)$ is finite at $\bar v=\qs$, where the vertices reach the light quarks. So, to get further, we must consider the configuration shown in Figure \ref{c53} on the left. One can think of that as the  strings (5) and (6)  
\noindent collapsing to a point. Note that the single baryon vertex remains at $r_v=r_{\vs}$. In this case, the boundary conditions \eqref{boundary-s2} and action \eqref{actionIIc3} become

\begin{equation}\label{boundary-c3}
x^{(1;2)}(0)=\mp\oh\ell\,,
\qquad
x^{(3)}(r_{\vs})=x^{(3)}(\rq)=-x_v
\,.
\qquad
x^{(1,2)}(\rvb)=0\,
\end{equation}
and 
\begin{equation}\label{Sc31}
S=\g T
\biggl(
2 \int_{0}^{\rvb} \frac{dr}{r^2}\,\ep^{\s r^2}\sqrt{1+(\partial_r x)^2}
+
\int_{r_{\vs}}^{\rq} \frac{dr}{r^2}\,\ep^{\s r^2}
+
3\k\,\frac{\ep^{-2\s r_{\vs}^2}}{r_{\vs}}
+
\frac{2}{\rvb}
\Bigl(
3\k\,\ep^{-2\s\rvb^2}
+
\n\ep^{\frac{1}{2}\s\rvb^2}
\Bigr)
+
\n\frac{\ep^{\frac{1}{2}\s\rq^2}}{\rq}
\,\biggr)
\,.
\end{equation}
Varying the action with respect to $\rq$ leads to Eqs.\eqref{q}, as before. However, varying the action with respect to $\rvb$ leads to

\begin{equation}\label{alphac3}
\sin\alpha=3\k(1+4\bar v)\ep^{-3\bar v}+\n(1-\bar v)\ep^{-\oh \bar v}
\,.
\end{equation}

Since the tangent angle $\alpha$ is non-negative, the formula \eqref{lc-s} for the separation distance still holds. On the other hand, the formula \eqref{Ec-s2} for the energy of the configuration is replaced by 

\begin{equation}\label{E-c3}
E_{\QQbqqq}
=2\g\sqrt{\s}
\biggl(
{\cal E}^+(\alpha,\bar v)
+\frac{3\k\ep^{-2\bar v}+\n\ep^{\oh\bar v}}{\sqrt{\bar v}}
\biggr)
+
E_0
+
2c
\,.
\end{equation}

For the parameter values we are using, $\alpha$ is a decreasing function of $\bar v$. It reaches zero at $\bar v = \Vz$, which is a solution to the equation

\begin{equation}\label{v0}
3\k(1+4\bar v)+\n(1-\bar v)\ep^{\frac{5}{2} \bar v}=0
\,.
\end{equation}
This solution defines the upper limit for $\bar v$. Therefore, the energy of the configuration is given in parametric form by $E_{\QQbqqq}=E_{\QQbqqq}(\bar v)$ and $\ell=\ell(\bar v)$, with the parameter varying from $\qs$ to $\Vz$.

This is not the whole story, however, as $\ell$ remains finite at $\bar v=\Vz$. So, we come to the question of what to do about it. The answer is that if $\alpha$ changes sign from positive to negative, $\ell$ continues to increase. In this case, the configuration profile becomes convex near $x=0$, as shown in Figure \ref{c53} on the right. The strings continue to descend deeper in the bulk until they finally reach the soft wall. As a result, the separation between the heavy quark sources becomes infinite.  

The expressions for the separation distance and energy can be obtained by simply replacing ${\cal L}^+$ and ${\cal E}^+$ with ${\cal L}^-$ and ${\cal E}^-$, as explained in Appendix B of \cite{a-stb3q}. So, we have 

\begin{equation}\label{l-c4}
\ell=\frac{2}{\sqrt{\s}}{\cal L}^-(\lambda,\bar v)
\,
\end{equation}
and 
\begin{equation}\label{E-c4}
E_{\QQbqqq}
=2\g\sqrt{\s}
\biggl(
{\cal E}^-(\lambda,\bar v)
+\frac{3\k\ep^{-2\bar v}+\n\ep^{\oh\bar v}}{\sqrt{\bar v}}
\biggr)
+
E_0
+
2c
\,.
\end{equation}
The functions ${\cal L}^-$ and ${\cal E}^-$ are given in Appendix A. The dimensionless parameter $\lambda$ is defined by $\lambda=\s\r0^2$, where $\r0=\max r(x)$ (see Figure \ref{c53}). Using \eqref{alphac3}, $\lambda$ can be conveniently expressed in terms of $\bar v$ as \cite{a-stb3q}

\begin{equation}\label{lambdav}
\lambda(\bar v)=-\text{ProductLog}\biggl[-\bar v\ep^{-\bar v}
\biggl(1-\Bigl(3\k(1+4\bar v)\ep^{-3\bar v}+\n(1-\bar v)\ep^{-\oh\bar v}\Bigr)^2\biggr)^{-\oh}
\biggr]
\,.	
\end{equation}
Here $\text{ProductLog}(z)$ denotes the principal solution for $w$ in $z=w\ep^{w}$ \cite{wolfram}. 

The parameter $\bar v$ varies from $\Vz$ to $\Vo$, which is found by solving the equation $\lambda=1$, or equivalently the equation 

\begin{equation}\label{v1}
\sqrt{1-\bar v^2\ep^{2(1-\bar v)}}+3\k(1+4\bar v)\ep^{-3\bar v}+\n(1-\bar v)\ep^{-\oh \bar v}=0
\,.	
\end{equation}
This is because ${\cal L}^-$ becomes infinite at $\lambda=1$ (see Appendix A).

To summarize, $E_{\QQbqqq}$ is a piecewise function of $\ell$, and the shape of the configuration (c) depends on the separation distance between the heavy quark sources. Furthermore, for $\ell>\ell(\vs)$ the model provides an explicit realization of the antiquark-diquark-diquark scheme of the pentaquark, as proposed in \cite{maiani}.

\subsubsection{The limiting cases}

As preparation for computing critical separations, we need some details on the behavior of $E_{\QQbqqq}$ for both small and large $\ell$. We begin with the case of small $\ell$. The relevant configuration for such a limit is depicted in Figure \ref{c51}, because $\ell$ vanishes at $\bar v=0$. So, taking the limit $\bar v\rightarrow 0$ in Eqs.\eqref{lc-s} and \eqref{Ec-s} with the help of Eqs.\eqref{fL+smallx} and \eqref{fE+smallx}, we find 

\begin{equation}\label{small}
\ell=\sqrt{\frac{\bar v}{\s}}\Bigl(\ell_0+\ell_1\bar v\Bigr)+o\bigl(\bar v^{\frac{3}{2}}\bigr)
\,,\qquad
E_{\QQbqqq}=\g\sqrt{\frac{\s}{\bar v}}\Bigl(E_0+E_1\bar v\Bigr)+E_{\Qqq}+E_{\Qqb}+o\bigl(\bar v^{\frac{1}{2}}\bigr)
\,.
\end{equation}
The expansion coefficients are given by  
\begin{gather}\label{lE}
\ell_0=\frac{1}{2}\tau^{-\frac{1}{2}}B\bigl(\tau^2;\tfrac{3}{4},\tfrac{1}{2}\bigr)
\,,
\qquad
\ell_1=\oh\tau^{-\frac{3}{2}}
\Bigl(3\tau\frac{1+2\k}{2+3\k}B\bigl(\tau^2;\tfrac{3}{4},-\tfrac{1}{2}\bigr)
-
B\bigl(\tau^2;\tfrac{5}{4},-\tfrac{1}{2}\bigr)
\Bigr)\,,
\\
E_0=2(1+3\k)+\oh\tau^{\oh}B\bigl(\tau^2;-\tfrac{1}{4},\tfrac{1}{2}\bigr)
\,,\quad
E_1=\frac{3}{2}(1+2\k)\Bigl(
-\frac{12\k}{1+3\k}
+\frac{\tau^{\oh}}{2+3\k}B\bigl(\tau^2;\tfrac{3}{4},-\tfrac{1}{2}\bigr)
-\frac{\tau^{-\oh}}{1+2\k}B\bigl(\tau^2;\tfrac{5}{4},-\tfrac{1}{2}\bigr)
\Bigr)
\,,
\end{gather}
where $\tau=\sqrt{-3\k(2+3\k)}\,$.\footnote{Note that $\tau$ is real with our choice of $\k=\kv$ (see the next subsection).} It is easy to eliminate $\bar v$ from the pair of equations \eqref{small}
 to obtain a nonlinear expression for $E_{\QQbqqq}$

\begin{equation}\label{EQQbqqq-small}
E_{\QQbqqq}=-\frac{\alpha_{\QQbqqq}}{\ell}+E_{\Qqq}+E_{\Qqb}+\boldsymbol{\sigma}_{\QQbqqq}\,\ell+
o(\ell)
\,.
\end{equation}
Here
\begin{equation}\label{alpha-sigmaUV}
\alpha_{\QQbqqq}=-\g \ell_0 E_0\,,
\qquad
\boldsymbol{\sigma}_{\QQbqqq}=\frac{1}{\ell_0}\Bigl(E_1+\frac{\ell_1}{\ell_0}E_0\Bigr)\g\s
\,.
\end{equation}
Interestingly, the leading term in \eqref{EQQbqqq-small} is the same as in the small-$\ell$ expansion of $E_{\QQbqqb}$ which is the energy of the connected tetraquark configuration for the $Q\bar Qq\bar q$ system \cite{a-QQbqqb}. This has an intuitive explanation: the presence of one additional light quark has no impact on the behavior of the heavy quark sources at extremely small separations. 

Moving on to the case of large $\ell$, we can use the relation \eqref{maiani} along with the asymptotic expansion of $E_{\QbQbqq}$ \cite{a-QQqbqb} to derive the corresponding expression for $E_{\QQbqqq}$. So, we get

\begin{equation}\label{EQQbqqq-large}
	E_{\QQbqqq}=\sigma\ell-2\g\sqrt{\s}\,I_{\QQbqqq}+E_0+2c+o(1)
	\,,
\qquad
\text{with}
\qquad
I_{\QQbqqq}={\cal I}(\Vo)
-
\frac{\n\ep^{\oh \Vo}+3\k\ep^{-2\Vo}}{\sqrt{\Vo}}
\,.
\end{equation}
The function ${\cal I}$ is defined in Appendix A. Notably, the constant term in the above expansions differs from each other. On the other hand, the coefficient $\sigma$ remains the same in all the known cases of connected string configurations ($Q\bar Q$ \cite{az1}, $QQQ$ \cite{a-3qPRD}, $QQq$ \cite{a-QQq}, etc.), as expected for the string tension.

Another useful expansion for $E_{\QQbqqq}$ is as follows. Instead of directly using the expressions \eqref{lc-s} and \eqref{Ec-s}, we could take the relation \eqref{maiani} and formally expand $E_{\QbQbqq}(\ell)$ in powers of $\ell$. This results in
\begin{equation}\label{approx0.2}
E_{\QQbqqq}
=
-\frac{\alpha_{\QQ}}{\ell}+2E_{\Qqq}-E_{\Qqb}+\boldsymbol{\sigma}_{\QQ}\,\ell+c+
o(\ell)
\,.
\end{equation}
We have used  the small-$\ell$ expansion of $E_{\QQqbqb}$ \cite{a-QQqbqb} together with $E_{\QbQbqq}=E_{\QQqbqb}$. The coefficients $\alpha_{\QQ}$ and $\boldsymbol{\sigma}_{\QQ}$ are defined similarly to $\alpha_{\QQbqqq}$ and $\boldsymbol{\sigma}_{\QQbqqq}$, but with the $\ell$'s and $E$'s replaced by 

\begin{gather}\label{lEapprox}
\boldsymbol{\ell}_0=\frac{1}{2}\xi^{-\frac{1}{2}}B\bigl(\xi^2;\tfrac{3}{4},\tfrac{1}{2}\bigr)
\,,
\qquad
\boldsymbol{\ell}_1=\oh\xi^{-\frac{3}{2}}
\Bigl(\bigl(2\xi+\frac{3}{4}\frac{\k-1}{\xi}\bigr)
B\bigl(\xi^2;\tfrac{3}{4},-\tfrac{1}{2}\bigr)
-
B\bigl(\xi^2;\tfrac{5}{4},-\tfrac{1}{2}\bigr)
\Bigr)\,,
\\
\boldsymbol{E}_0=1+3\k+\oh\xi^{\oh}B\bigl(\xi^2;-\tfrac{1}{4},\tfrac{1}{2}\bigr)
\,,\quad
\boldsymbol{E}_1=\xi\boldsymbol{\ell}_1-1-6\k+\oh B\bigl(\xi^2;\tfrac{1}{4},\tfrac{1}{2}\bigr)
\,,
\end{gather}
where $\xi=\frac{\sqrt{3}}{2}\sqrt{1-2\k-3\k^2}$. This approximation has an advantage over the expansion \eqref{EQQbqqq-small} near $\ell=0.2\,\text{fm}$, as we will see shortly. 

\subsubsection{Putting all the pieces together}

Now let's discuss the gluing of all the branches of $E_{\QQbqqq}(\ell)$. For this, we need to specify the model parameters. Here, we use one of the two parameter sets suggested in \cite{a-strb}, which is mainly resulted from fitting the lattice QCD data to the string model we are considering. The value of $\s$ is fixed from the slope of the Regge trajectory of $\rho(n)$ mesons in the soft wall model with the geometry \eqref{metric}. As a result, we get $\s=0.45\,\text{GeV}^2$ \cite{a-q2}. Then, fitting the value of the string tension $\sigma$ (see Eq.\eqref{EQQb-large}) to its value in \cite{bulava}, we get $\g=0.176$. The parameter $\n$ is adjusted to reproduce the lattice result for the string breaking distance in the $Q\bar Q$ system. With $\boldsymbol{\ell}_{\QQb}=1.22\,\text{fm}$ for the $u$ and $d$ quarks \cite{bulava}, we get $\n=3.057$ \cite{a-strb}. 

  In principle, the value of $\k$ could be adjusted to fit the lattice data for the three-quark potential, as done in \cite{a-3qPRD} for pure $SU(3)$ gauge theory. But there are no lattice data available for QCD with two light quarks. There are still two special options: $\k=-0.102$ motivated by phenomenology\footnote{Note that $\k=-0.102$ is a solution to the equation $\alpha_{\QQ}(\k)=\oh\alpha_{\QQb}$, which follows from the phenomenological rule $E_{\QQ}(\ell)=\oh E_{\QQb}(\ell)$ in the limit $\ell\rightarrow 0$.} and $\k=-0.087$ obtained from the lattice data for pure gauge theory \cite{a-3qPRD}. However, both values are outside of the range of allowed values for $\k$ as follows from the analysis of Eq.\eqref{v}. Therefore, in this situation, it is reasonable to choose $\k=\kv$ with is the closest to those values.

Having fixed the model parameters, we can immediately perform some simple but important calculations. First, let's check that $\qs>\vs$. That is, our construction of the string configurations makes sense. From Eqs.\eqref{q} and \eqref{v}, we find that $\qs=0.566$ and $\vs=\frac{1}{12}$, as desired. In addition, from \eqref{v0} and \eqref{v1}, we get $\Vz=0.829$ and $\Vo=0.930$. Second, given $\vs$, one can immediately estimate the smallest separation between the heavy quarks for the configuration shown in Figure \ref{c52}(c). This gives $\ell(\vs)=0.106\,\text{fm}$. It is quite surprising that the antiquark-diquark-diquark scheme arises already at such small separations.

\begin{figure}[H]
\centering
\includegraphics[width=7.25cm]{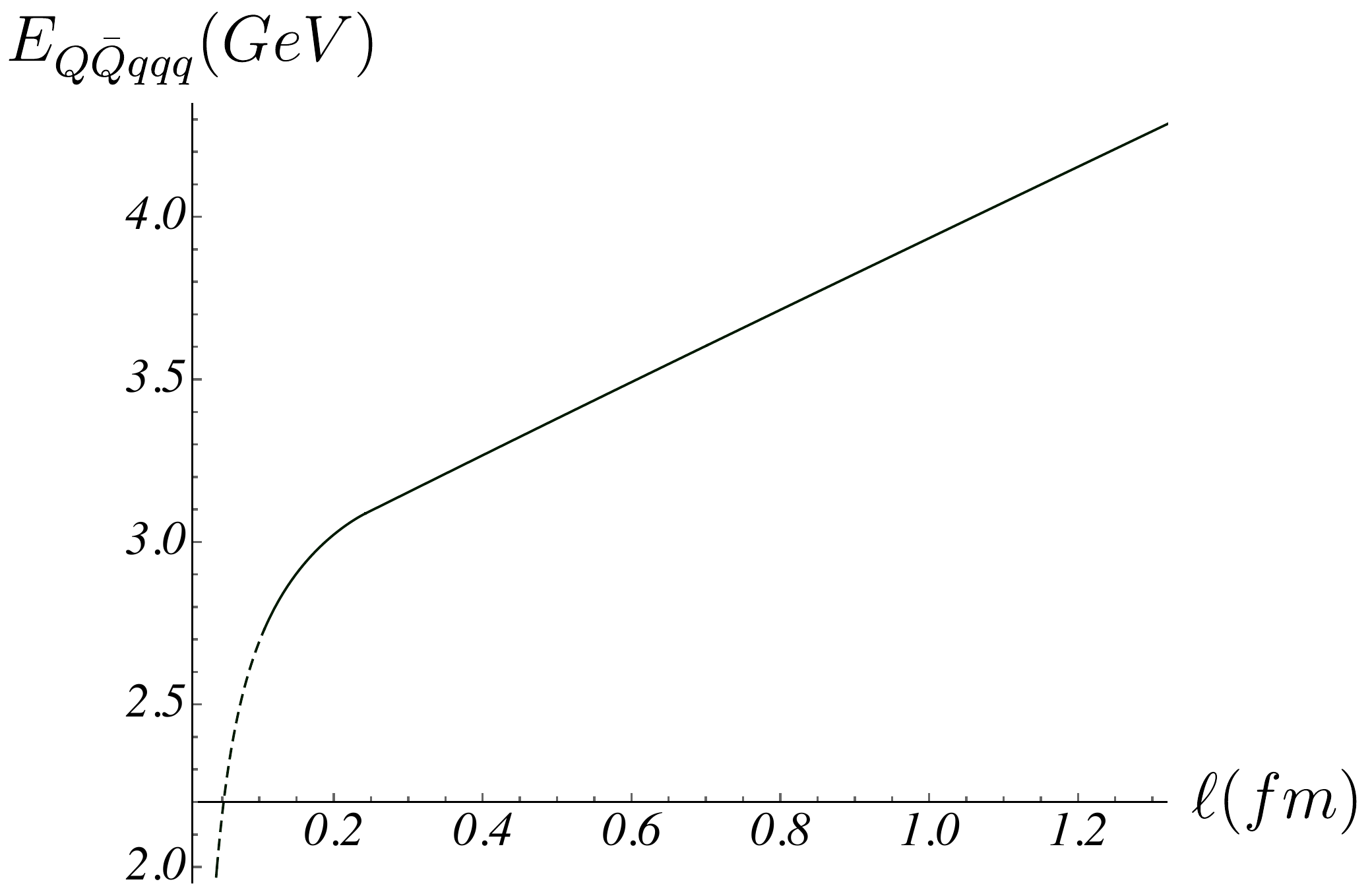}
\hspace{2.5cm}
\includegraphics[width=6.75cm]{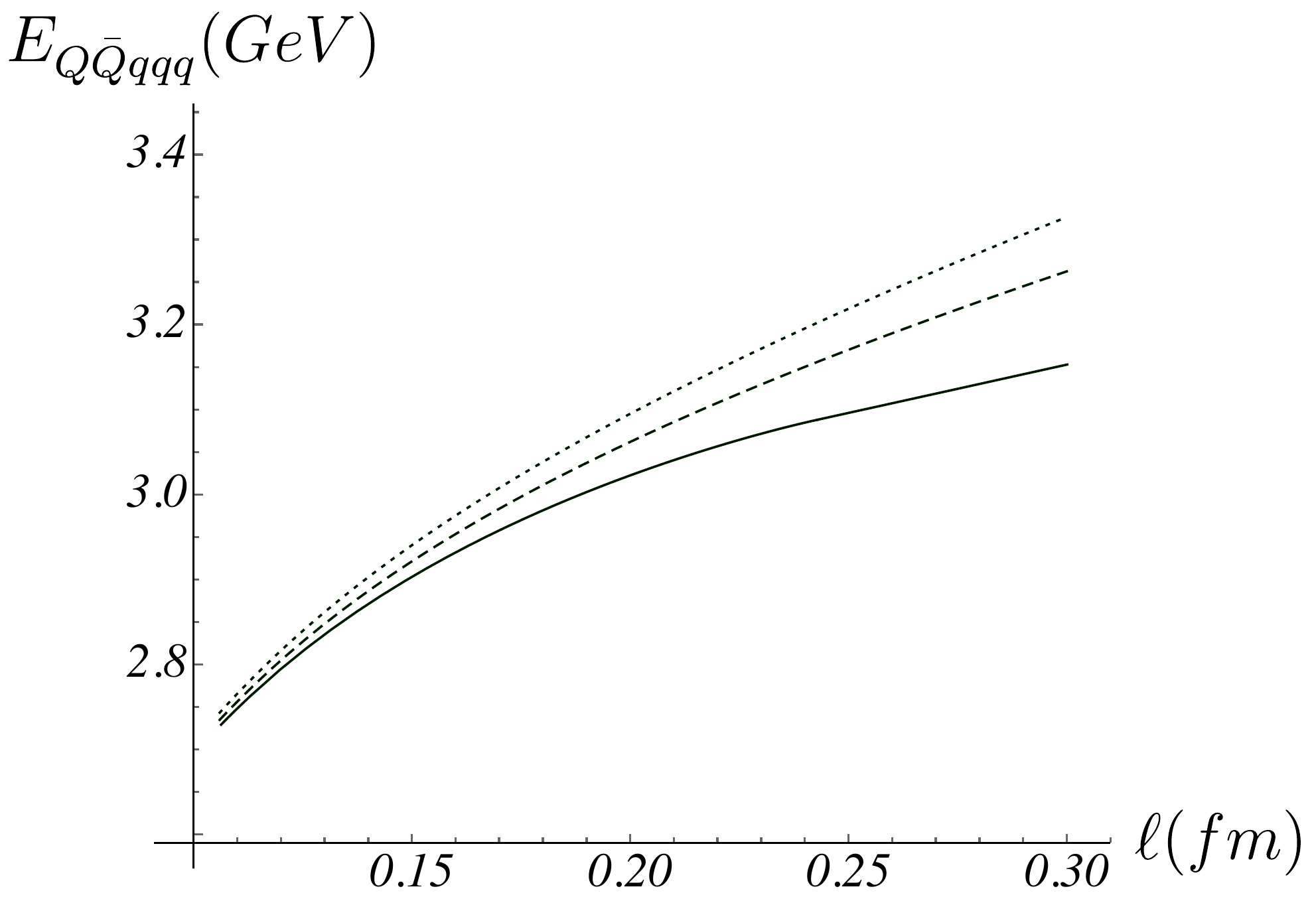}
\caption{{\small $E_{\QQbqqq}$ vs $\ell$. In this and subsequent Figures, we set $c=0.623\,\text{GeV}$. Left: The result of plotting the piecewise function $E_{\QQbqqq}$. The dashed curve corresponds to the configuration shown in Figure \ref{c51}, while the solid curve to the remaining configurations for which the antiquark-diquark-diquark scheme holds.  Right: The function near $\ell=0.20\,\text{fm}$. The dotted and dashed curves correspond to the approximations \eqref{EQQbqqq-small} and \eqref{approx0.2}, respectively.}}
\label{Plotc}
\end{figure}

Plotting $E_{\QQbqqq}$ as a function of $\ell$ has become a straightforward task. The result is shown in Figure \ref{Plotc} on the left. 
\noindent From this Figure it is seen that all the pieces of the function are smoothly glued together. Additionally, $E_{\QQbqqq}(\ell)$ approximates to a linear function for separations greater than $0.45,\text{fm}$. For future reference, we note that the function $E_{\QQbqqq}(\ell)$ is better approximated by \eqref{approx0.2} than by \eqref{EQQbqqq-small} near $\ell=0.20\,\text{fm}$. This is illustrated in the above Figure on the right.  

\subsection{The disconnected configurations (d)-(f)}

\begin{figure}[htbp]
\centering
\includegraphics[width=5.1cm]{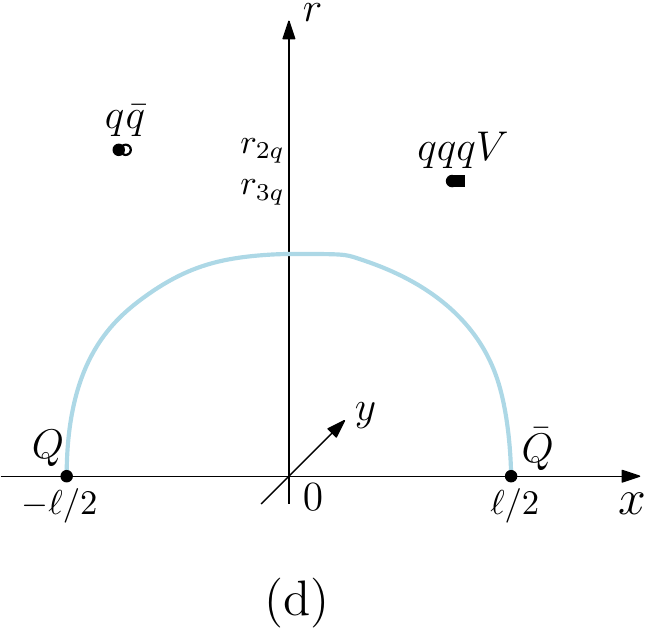}
\hspace{1.12cm}
\includegraphics[width=5.1cm]{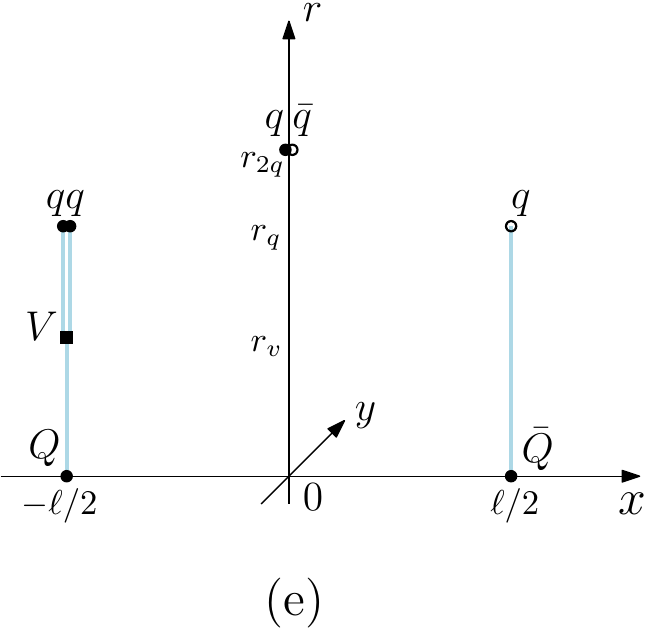}
\hspace{1.12cm}
\includegraphics[width=5.1cm]{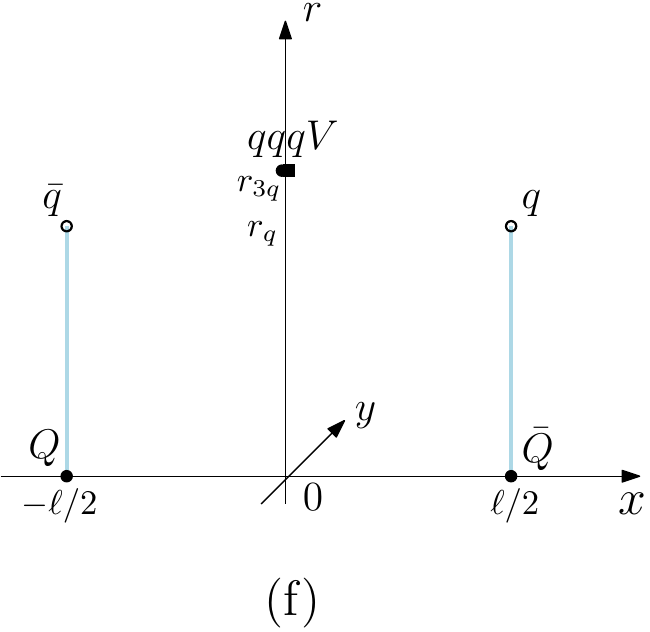}
\caption{{\small Configurations (d), (e), and (f) in five dimensions.}}
\label{con-def}
\end{figure}

We begin by considering configuration (d), which is obtained by adding a $q\bar q$ pair (pion) to configuration (a). The pion is placed in the interior at $r=r_{\text{\tiny 2q}}$, resulting in the configuration shown in Figure \ref{con-def}(d). This configuration can 
be interpreted as a hadroquarkonium state: a $Q\bar Q$ pair in a pion-nucleon cloud. Although there are no calculations available for this case on the lattice, we will assume that adding a pion and averaging over its position leads to an energy increase by $E_{\qqb}$. Thus, the total energy is 

\begin{equation}\label{Ed}
E^{\text{(d)}}=E^{\text(a)}+E_{\qqb}=E_{\QQb}+E_{\nucl}+E_{\qqb}
\,.
\end{equation}
$E_{\qqb}$ was computed in \cite{a-QQbqqb} with the result

\begin{equation}\label{Eqqb}
E_{\qqb}=2\n\sqrt{\g\sigma}
\,.
\end{equation}
Here $\sigma$ is the string tension. Note that $r_{\nucl}<r_{\text{\tiny 2q}}=1/\sqrt{\s}$.

Similarly, configuration (e) is obtained by adding a $q\bar q$ pair to configuration (b). In five dimensions, the corresponding configuration is shown in Figure \ref{con-def} (e). It can be interpreted as a pair of heavy-light hadrons in a pion cloud. By the same assumption that we have made previously in our treatment of configuration (d), the energy is given by  

\begin{equation}\label{Ee}
E^{\text{(e)}}=E^{\text(b)}+E_{\qqb}=E_{\Qqb}+E_{\Qqq}+E_{\qqb}
\,.
\end{equation}

Finally, for configuration (f) we expect 

\begin{equation}\label{Ef}
E^{\text{(f)}}=2E_{\Qqb}+E_{\nucl}
\,,
\end{equation}
as in \cite{a-QQ3q}. This configuration can be interpreted as a pair of heavy-light mesons surrounded by a nucleon cloud. It is worth noting that it may arise from configuration (a) through string breaking in the $Q\bar Q$ pair (see Figure \ref{5QQb}).

\subsection{What we have learned}

\begin{figure}[htbp]
\centering
\includegraphics[width=9cm]{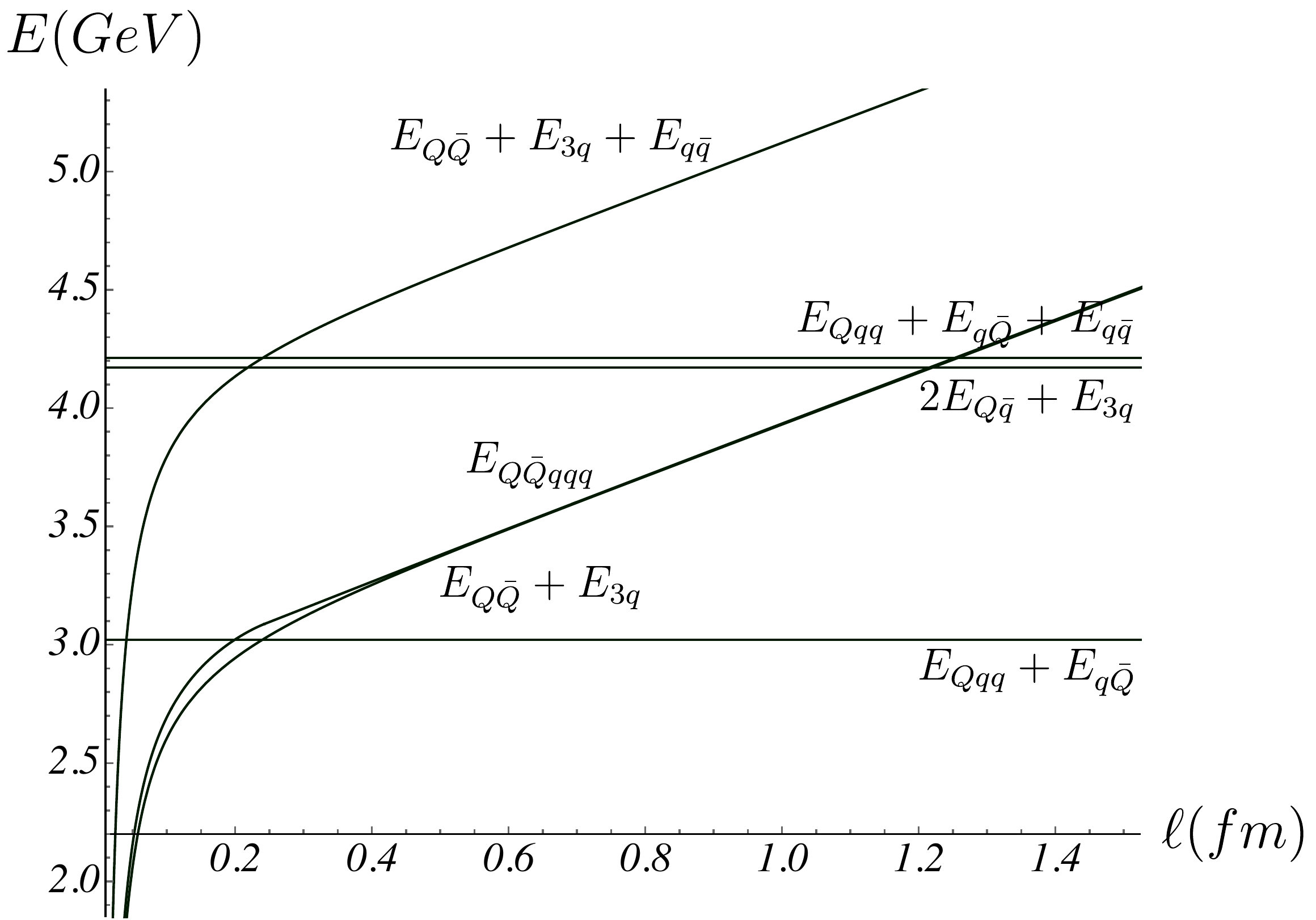}
\caption{{\small Various $E$ vs $\ell$ plots. Here we assume that $E_{\Qqb}=E_{\qQb}$.}}
\label{all-L}
\end{figure}

It is instructive to see how the energies of the configurations mentioned above depend on the separation between the heavy quark-antiquark pair. In Figure \ref{all-L} we plot those for our parameter values. It is obvious from the plot that 
\noindent the energy of the ground state is determined by the contributions from configurations (a) and (b). Therefore, we have $V_0=\min\{E_{\QQb}+E_{\nucl}, E_{\Qqq}+E_{\qQb}\}$. The potential interpolates between $E_{\QQb}+E_{\nucl}$ at small separations and $E_{\Qqq}+E_{\qQb}$ at larger ones. An important observation is that the transition between these two regimes occurs at a relatively small length scale of about $0.2\,\text{fm}$. To quantify this observation, we define a critical separation distance by 

\begin{equation}\label{lQq}
E_{\QQb}(l_{\Qq})+E_{\nucl}=E_{\Qqq}+E_{\Qqb}
\,.
\end{equation}
It is natural to interpret $l_{\Qq}$ as a scale that distinguishes the descriptions in terms of the hadroquarkonium state and hadronic molecule. From a string theory perspective, the transition occurs through string reconnection: $Q\bar Q+qqq\rightarrow Qqq+q\bar Q$, as sketched in Figure \ref{sint}(a).  

Due to small value of $l_{\Qq}$, we can solve the equation approximately by neglecting all but the first three terms in $E_{\QQb}$. With the help of \eqref{EQQb-small}, the solution can be witten as  

\begin{equation}\label{srec-small}
l_{\Qq}\approx
\frac{1}{2\boldsymbol{\sigma}_{\QQb}}
\Bigl(E_{\Qqq}+E_{\Qqb}-E_{\nucl}-2c\Bigr)+
\sqrt{\frac{\alpha_{\QQb}}{\boldsymbol{\sigma}_{\QQb}}
+
\frac{1}{4\boldsymbol{\sigma}^2_{\QQb}}
\Bigl(E_{\Qqq}+E_{\Qqb}-E_{\nucl}-2c\Bigr)^2}
\,.
\end{equation}
An important fact is that the critical separation distance is independent of $c$, as follows from the expressions for $E_{\Qqq}$ and $E_{\Qqb}$. Let's make a simple estimate of $l_{\Qq}$. For our chosen parameter values, we have  

\begin{equation}\label{lcnum}
l_{\Qq}\approx0.241\,\text{fm}
\,.
\end{equation}
Thus, this simple estimate suggests that $l_{\Qq}$ is indeed of order $0.2\,\text{fm}$, as expected. 
 
Before proceeding further, we discuss here a point with the plots. As seen from the Figure, for separations greater than about $0.4,\text{fm}$, the difference between the plots for $E_{\QQb}+E_{\nucl}$ and $E_{\QQbqqq}$ becomes negligible. To see which configuration, (a) or (c), has a higher energy, we can compare their behavior for large $\ell$. Using \eqref{EQQbqqq-large} and \eqref{EQQb-large}, we get  

\begin{equation}\label{Gap}
\Delta=E_{\QQbqqq}-E_{\QQb}-E_{\nucl}=
2\g\sqrt{\s}\bigl(I_0-I_{\QQbqqq}\bigr)-E_{\nucl}
\,. 
\end{equation}
Combining this with \eqref{nucl} yields $\Delta\approx 6,\text{MeV}$ for our parameter values, indicating that configuration (c) has a higher energy than configuration (a). 

With this in mind, we can formally define the B-O potential for the first excited state as $V_1=\min\{E_{\QQbqqq},E_{\Qqq}+E_{\qQb},E_{\QQb}+E_{\nucl}, 2E_{\Qqb}+E_{\nucl}\}$. This definition leads to the emergence of three distinct scales that separate different configurations, or in other words different descriptions. The first is a scale which refers to the process of string junction annihilation: $Q\bar Qqqq\rightarrow Qqq+q\bar Q$. In this case, we define a critical separation distance by 

\begin{equation}\label{sja}
	E_{\QQbqqq}(\boldsymbol{\ell}_{\QQbqqq})=E_{\Qqq}+E_{\Qqb}
	\,.
	\end{equation}
The scale $\boldsymbol{\ell}_{\QQbqqq}$, with a value of about $0.2,\text{fm}$, distinguishes the descriptions in terms of the compact pentaquark state and hadronic molecule. Using the asymptotic approximation \eqref{approx0.2}, we find 
\begin{equation}\label{sja2}
	\boldsymbol{\ell}_{\QQbqqq}\approx
	\frac{1}{2\boldsymbol{\sigma}_{\QQ}}
\Bigl(2E_{\Qqb}-E_{\Qqq}-c\Bigr)+
\sqrt{\frac{\alpha_{\QQ}}{\boldsymbol{\sigma}_{\QQ}}
+
\frac{1}{4\boldsymbol{\sigma}^2_{\QQ}}
\Bigl(2E_{\Qqb}-E_{\Qqq}-c\Bigr)^2}
\,.
\end{equation}
The same argument that we have already given for $l_{\Qq}$ shows that $\boldsymbol{\ell}_{\QQbqqq}$ is independent of $c$.  Note that the above expression is identical to that obtained in \cite{a-QQqbqb} for the $QQ\bar q\bar q$ system.\footnote{This is due to our use of the approximation \eqref{approx0.2}.} Because of this, the same estimate gives 

\begin{equation}\label{sja-es}
\boldsymbol{\ell}_{\QQbqqq}\approx 0.184\,\text{fm}
\,.
\end{equation}

The second scale is related to the process of string reconnection, which we have just discussed above, but now in the opposite direction: $Qqq+q\bar Q\rightarrow Q\bar Q+3q$. Therefore the formula \eqref{srec-small} for the critical separation distance holds true. Finally, the third scale arises from string breaking: $Q\bar Q+3q\rightarrow Q\bar q+q\bar Q+3q$. Here we assume that a nucleon cloud has little impact on this process and, as a consequence, the formula \eqref{LQQb-large} for the string breaking distance $\ell_{\QQb}$ remains valid. If so, then $\ell_{\QQb}=1.22\,\text{fm}$ (for more on this point, see Appendix B). 

\section{Other elementary excitations}
\renewcommand{\theequation}{4.\arabic{equation}}
\setcounter{equation}{0}

\subsection{Preview}

\begin{figure}[H]
\centering
\includegraphics[width=14cm]{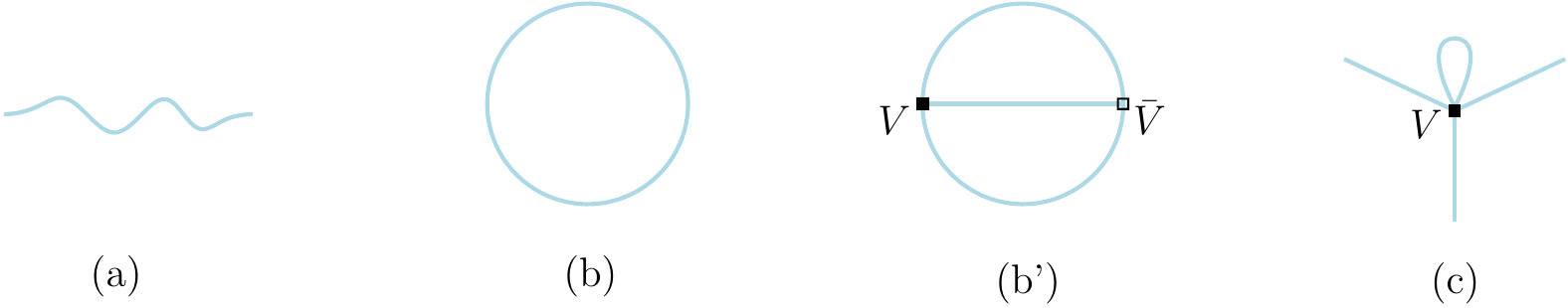}
\caption{{\small Some types of gluonic excitations.}}
\label{excitations}
\end{figure}

The assumptions made about excited states in Section II are oversimplified for higher-lying B-O potentials. For instance, when constructing configurations for excited states, one must consider excited strings such as the one depicted in Figure \ref{excitations}(a). These strings represent a type of gluonic excitations that has been studied in lattice QCD, but only
 \noindent  within the $Q\bar Q$ system \cite{swanson}. In the context of the current model, one of these excitations (type $\Sigma_u^-$) was modeled in \cite{a-hyb}, and later, it was considered within the $Q\bar Qq\bar q$ system in \cite{a-QQbqqb}. Additionally, glueballs must be included as another kind of gluonic excitations \cite{glueball}. Two of the simplest examples of such color-singlet states are sketched in Figures \ref{excitations}(b) and (b'). The former represents a closed string, while the latter involves a pair of baryon vertices connected by open strings. These gluonic excitations are natural from the perspective of string theory in four dimensions \cite{XA}.  However, in ten dimensions, there is a novelty related to the description of the baryon vertex as a five-brane \cite{witten}, which means that we must also consider brane excitations. This would give rise to a set of excited vertices that represent a new type of gluonic excitations. The simplest example of such an excitation is illustrated in Figure \ref{excitations}(c), where the excitation is due to an open string with endpoints on the brane.

\begin{figure}[htbp]
\centering
\includegraphics[width=8.5cm]{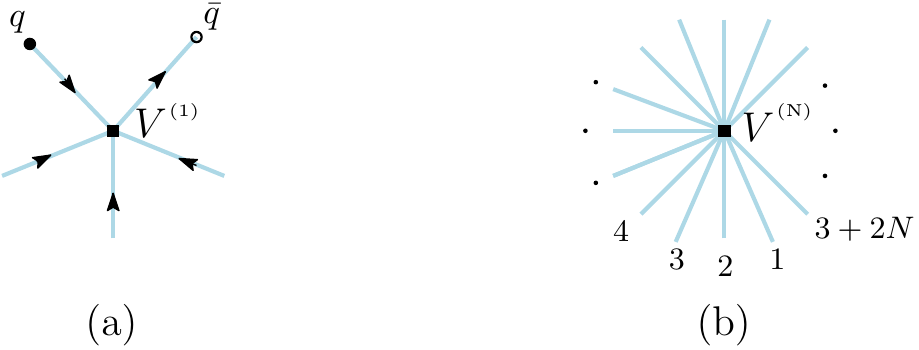}
\caption{{\small Generalized baryon vertices: $V^{(1)}$ (left) and $V^{(\text{N})}$ (right).}}
\label{h-vert}
\end{figure}

At this point, one might ask what happens when a string attached to the brane breaks due to the production of a $q\bar q$ pair. If so, this results in a simple picture shown in Figure \ref{h-vert}(a). If a string (a chromoelectric flux tube) goes from a quark  
\noindent to an antiquark, then the difference between the numbers of in- and out-strings is equal to $3$, which is precisely the number of colors. This example provides a natural definition of a baryon vertex $V^{(1)}$, where four in- and one out-strings meet. It is straightforward to suggest future generalizations and define a vertex $V^{(\text{N})}$ with $N+3$ in-strings and $N$ out-strings, as sketched in Figure \ref{h-vert}(b).\footnote{We have learned that Cobi Sonnenschein is also considering such vertices in QCD.} In this notation, the baryon vertex of Sec.II corresponds to $V^{(0)}$. 

\subsection{Implications for pentaquarks}

\begin{figure}[H]
\centering
\includegraphics[width=8.5cm]{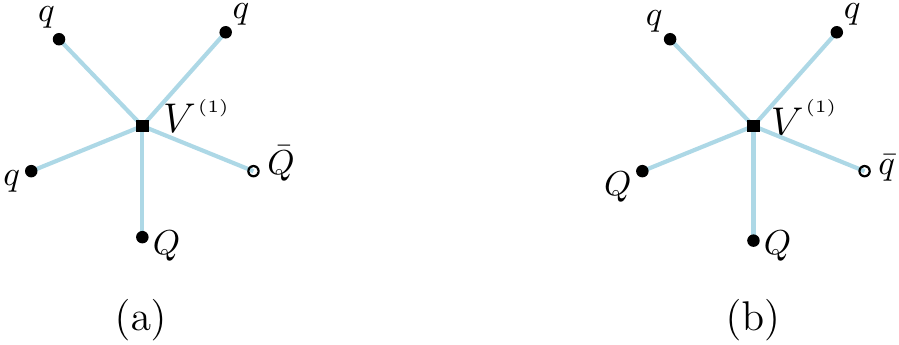}
\caption{{\small The generalized pentaquark configurations described by $V^{(1)}$ for the $Q\bar Qqqq$ and $QQqq\bar q$ systems.}}
\label{4h-penta}
\end{figure}

Finding evidence for generalized vertices in QCD, particularly for $V^{(1)}$, would be highly interesting. Perhaps the simplest way to achieve this is to examine hybrid potentials in the $QQQ$ quark system through lattice simulations. In the present context, $V^{(1)}$ produces an additional connected pentaquark configuration as that shown in Figure \ref{4h-penta}(a).\footnote{A similar configuration also occurs for the $QQqq\bar q$ system. See Figure \ref{4h-penta}(b).} 
\noindent Our aim for this subsection is to construct a five-dimensional counterpart of this configuration. 

\begin{figure}[htbp]
\centering
\includegraphics[width=6.5cm]{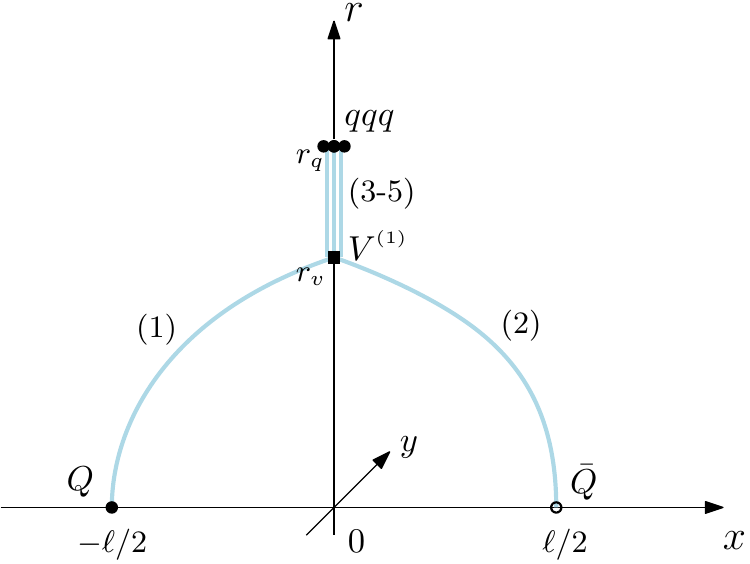}
\caption{{\small A possible configuration of Figure \ref{4h-penta}(a) in five dimensions.}}
\label{5hs}
\end{figure}

If one places the above configuration on the boundary of five-dimensional space, a gravitational force pulls the light quarks and strings towards the interior. As a result, the configuration takes the form shown in Figure \ref{5hs}, which
    is supposed to be the configuration of lowest energy due to its high degree of symmetry.  

The total action governing this configuration is 

\begin{equation}\label{Sh1}
S=\sum_{i=1}^5 S_{\text{\tiny NG}}^{(i)}+S_{\text{vert}}^{(1)}+3S_{\text q}
\,.
\end{equation}
For what follows, we will assume that the action for the vertex $V^{(1)}$ is also given by the five-brane world volume action \eqref{baryon-v}, specifically $S_{\text{vert}}^{(1)}=S_{\text{vert}}$. From this starting point, the analysis proceeds in an obvious manner. However, a quicker way to proceed is to use the results from Appendix C. Using those, we can get the corresponding formulas by rescaling $\k\rightarrow\tfrac{1}{3}\k$. So, we have 

\begin{equation}\label{Eh1}
E^{(1)}_{\QQbqqq}=3\g\sqrt{\s}
\biggl(
\frac{2}{3}{\cal E}^+(\alpha,v)
+
{\cal Q}(\qs)-{\cal Q}(v)
+
\k\frac{\ep^{-2v}}{\sqrt{v}}
+
\n\frac{\ep^{\oh \qs}}{\sqrt{\qs}}
\biggr)
+2c
\,,
\end{equation}
with the separation distance $\ell$ given by \eqref{lc-s}. The tangent angle $\alpha$ is expressed in terms of $v$ as

\begin{equation}\label{alphah1}
\sin\alpha=\frac{3}{2}\Bigl(1+\k(1+4v)\ep^{-3v}\Bigr)
\,.	
\end{equation}
The parameter $v$ ranges from $0$ to $\qs$. 

A simple analysis shows that for $\k=\kv$, Eq.\eqref{alphah1} has only one solution, which is $v=\vs$. In this case $\alpha=\pi/2$, $\ell=0$, and $E^{(1)}_{\QQbqqq}=3.022\,\text{GeV}$. Thus, the present configuration is subleading to the pentaquark configuration (c) of Sec.III. In fact, it may be spurious as it only exists at zero-separation of the quark-antiquark pair, where the string models are not reliable.  

\begin{figure}[H]
\centering
\includegraphics[width=6.5cm]{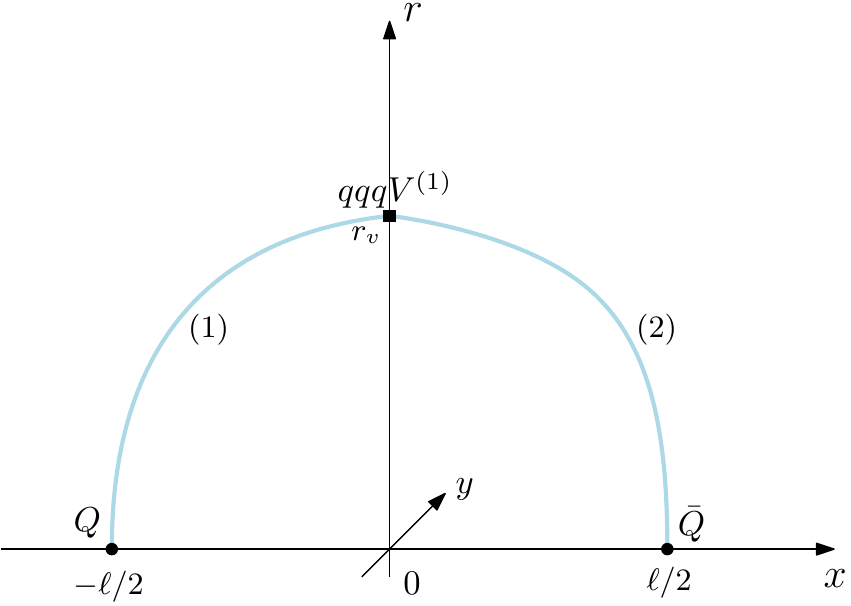}
\hspace{2cm}
\includegraphics[width=6.5cm]{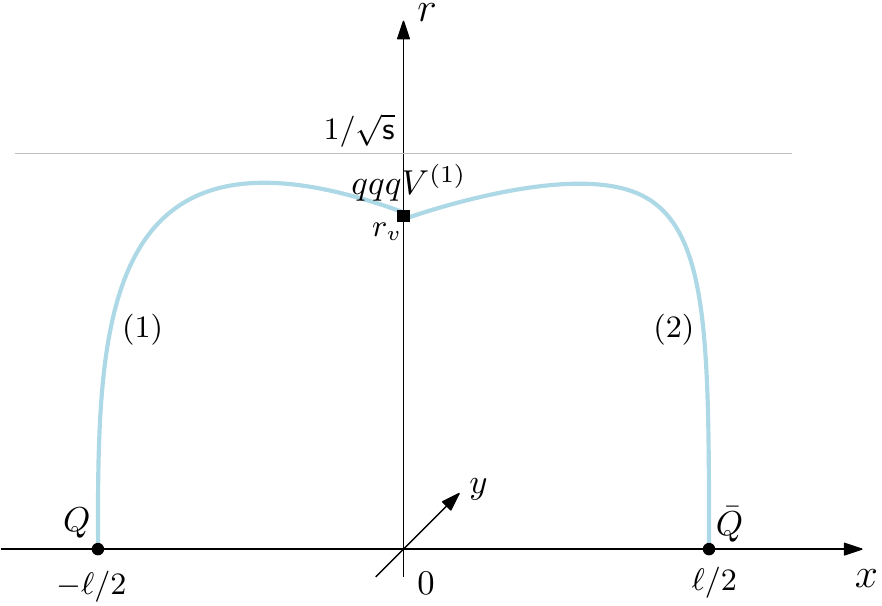}
\hspace{2cm}
\caption{{\small The generalized pentaquark configuration for small (left) and large (right) heavy quark separations.}}
\label{5hl}
\end{figure}

We can proceed further in the fashion just described in Sec.III. For $v$ slightly larger than $\qs$, the configuration transforms into the one shown in the left panel of Figure \ref{5hl}, where strings (3)-(5) collapse to a point. The remaining 
\noindent strings join in the interior that leads to the formation of cusp at $r=r_v$. For this case, the total action reduces to   
 
\begin{equation}\label{Sh2}
S=\sum_{i=1}^2 S_{\text{\tiny NG}}^{(i)}+S_{\text{vert}}+3S_{\text q}
\,.
\end{equation}

We take a shortcut to the desired results instead of directly analyzing \eqref{Sh2}. The expression for $\ell$ remains unchanged, and is given by \eqref{lc-s}, while the expression for the energy is obtained by setting $\qs=v$ in \eqref{Eh1}. So, we get

\begin{equation}\label{Eh2}
\ell=\frac{2}{\sqrt{\s}}{\cal L}^+(\alpha,v)
\,,
\qquad
E^{(1)}_{\QQbqqq}=3\g\sqrt{\s}
\biggl(
\frac{2}{3}{\cal E}^+(\alpha,v)
+
\frac{\k\ep^{-2v}+\n\ep^{\oh v}}{\sqrt{v}}
\biggr)
+2c
\,.
\end{equation}
The angle $\alpha$ is now determined from the equation 

\begin{equation}\label{alphah2}
\sin\alpha=\frac{3}{2}
\Bigl(\k(1+4v)\ep^{-3v}+\n(1-v)\ep^{-\oh v}\Bigr)
\,,
\end{equation}
which is obtained from Eq.\eqref{alphac3} by rescaling $\k\rightarrow \frac{1}{2}\k$ and $\n\rightarrow\frac{3}{2}\n$. An important fact about this equation is that the right-hand side, as a function of $v$, decreases as $v$ increases on the interval $[\qs,1]$, given the values of $\k$ and $\n$ set in Sec. III. It takes the value $1$ at $v=\ws$, where $\ws$ satisfies the equation

\begin{equation}\label{w0}
\k(1+4v)\ep^{-3v}+\n(1-v)\ep^{-\oh v}=\frac{2}{3}
	\,.
\end{equation}
The separation distance $\ell$ becomes zero at $v=\ws$ because $\cos\alpha=0$. Additionally, the right-hand side of \eqref{alphah2} is zero at $v=\wsz$ if $\wsz$ is a solution to the equation

\begin{equation}\label{w1}
\k(1+4v)\ep^{-3v}+\n(1-v)\ep^{-\oh v}=0
	\,. 
\end{equation}
At this point, the cusp disappears as $\cos\alpha=1$.  

If these solutions exist, the energy of the generalized pentaquark configuration can be expressed parametrically as $\ell=\ell(v)$ and $E^{(1)}_{\QQbqqq}=E^{(1)}_{\QQbqqq}(v)$, with the parameter taking values on the interval $[\ws,\wsz]$. 

Just as in Sec.III, to achieve an infinite separation distance, we must consider changing the sign of $\alpha$. This causes the configuration profile to become convex near $x=0$, as shown in Figure \ref{5hl} on the right. The strings reach the soft wall corresponding to an infinite separation for some value of $\alpha$. Since this modified configuration is governed by the action \eqref{Sh2}, we can obtain formulas for $\ell$ and $E^{(1)}_{\QQbqqq}$ by replacing ${\cal L}^+$ and ${\cal E}^+$ with ${\cal L}^-$ and ${\cal E}^-$. This results in 

\begin{equation}\label{Eh3}
\ell=\frac{2}{\sqrt{\s}}{\cal L}^-(\lambda,v)
\,,
\qquad
E^{(1)}_{\QQbqqq}=\g\sqrt{\s}
\biggl(
2{\cal E}^-(\lambda,v)
+
3\frac{\k\ep^{-2v}+\n\ep^{\oh v}}{\sqrt{v}}
\biggr)
+2c
\,.
\end{equation}
The function $\lambda$ is given by 

\begin{equation}\label{lambdah}
\lambda(v)=-\text{ProductLog}\biggl[-v\ep^{-v}
\biggl(1-\frac{9}{4}\Bigl(\k(1+4v)\ep^{-3v}+\n(1-v)\ep^{-\oh v}\Bigr)^2\biggr)^{-\oh}
\biggr]
\,,
\end{equation}
as follows from \eqref{alphah2} combined with the formula (B.18) in \cite{a-stb3q}.

The parameter $v$ now takes values on the interval $[\wsz,\wso]$, where the upper bound is determined from the equation $\lambda=1$, or equivalently the equation 

\begin{equation}\label{v1h}
2\sqrt{1-v^2\ep^{2(1-v)}}+3\k(1+4v)\ep^{-3v}+3\n(1-v)\ep^{-\oh v}=0
\,
\end{equation}
in the interval $[0,1]$. The reason for this is that ${\cal L}^-$ is singular at $\lambda=1$ (see Eq.\eqref{L-y=1}), resulting in an infinite separation distance between the heavy quark sources.

For future reference, let's briefly discuss the behavior of $E_{\QQbqqq}^{(1)}$ for small and large $\ell$. This may be done along the lines of subsection 4 of Sec.III, except that the right behavior for $\ell\rightarrow 0$ comes from the limit $v\rightarrow\ws$. 

When $\ell$ is small, we have

\begin{equation}\label{Eh-small}
	E^{(1)}_{\QQbqqq}=E^{(1)}_{\QQbqqq}(\ws)+A\ell^2+O(\ell^3)
	\,,
\end{equation}
with 
\begin{equation}
	E^{(1)}_{\QQbqqq}(\ws)=\g\sqrt{\s}\Bigl(2{\cal Q}(\ws)+3\frac{\k\ep^{-2\ws}+\n\ep^{\oh\ws}}{\sqrt{\ws}}\Bigr)+2c
	\,,\qquad
	A=\g\s^{\frac{3}{2}}\Bigl(\sqrt{\pi}\,\text{erf}(\sqrt{\ws})-2\sqrt{\ws}\,\ep^{-\ws}\Bigr)^{-1}
	\,.
\end{equation}
Here $\text{erf}(x)$ is the error function. 

On the other hand, for large $\ell$, $E_{\QQbqqq}^{(1)}$ behaves linearly with $\ell$. Explicitly,

\begin{equation}\label{Eh-large}
E^{(1)}_{\QQbqqq}(\ell)=\sigma\ell-2\g\sqrt{\s}\,I_{\QQbqqq}^{(1)}+2c+o(1)
\,,
\qquad\text{with}\qquad
I_{\QQbqqq}^{(1)}={\cal I}(\wso)
-
\frac{3}{2}\frac{\k\ep^{-2\wso}+\n\ep^{\oh\wso}}{\sqrt{\wso}}
\,
\end{equation}
and the same string tension $\sigma$  as in \eqref{EQQb-large}.
 
The analysis of the corresponding configuration for the $QQqq\bar q$ system in five dimensions is similar, except that we replace $\bar Q$ with $Q$ and one of the $q$'s with $\bar q$. At zero baryon chemical potential, the resulting formulas coincide with the formulas obtained above for the $Q\bar Qqqq$ system. 

\begin{figure}[H]
\centering
\includegraphics[width=8.15cm]{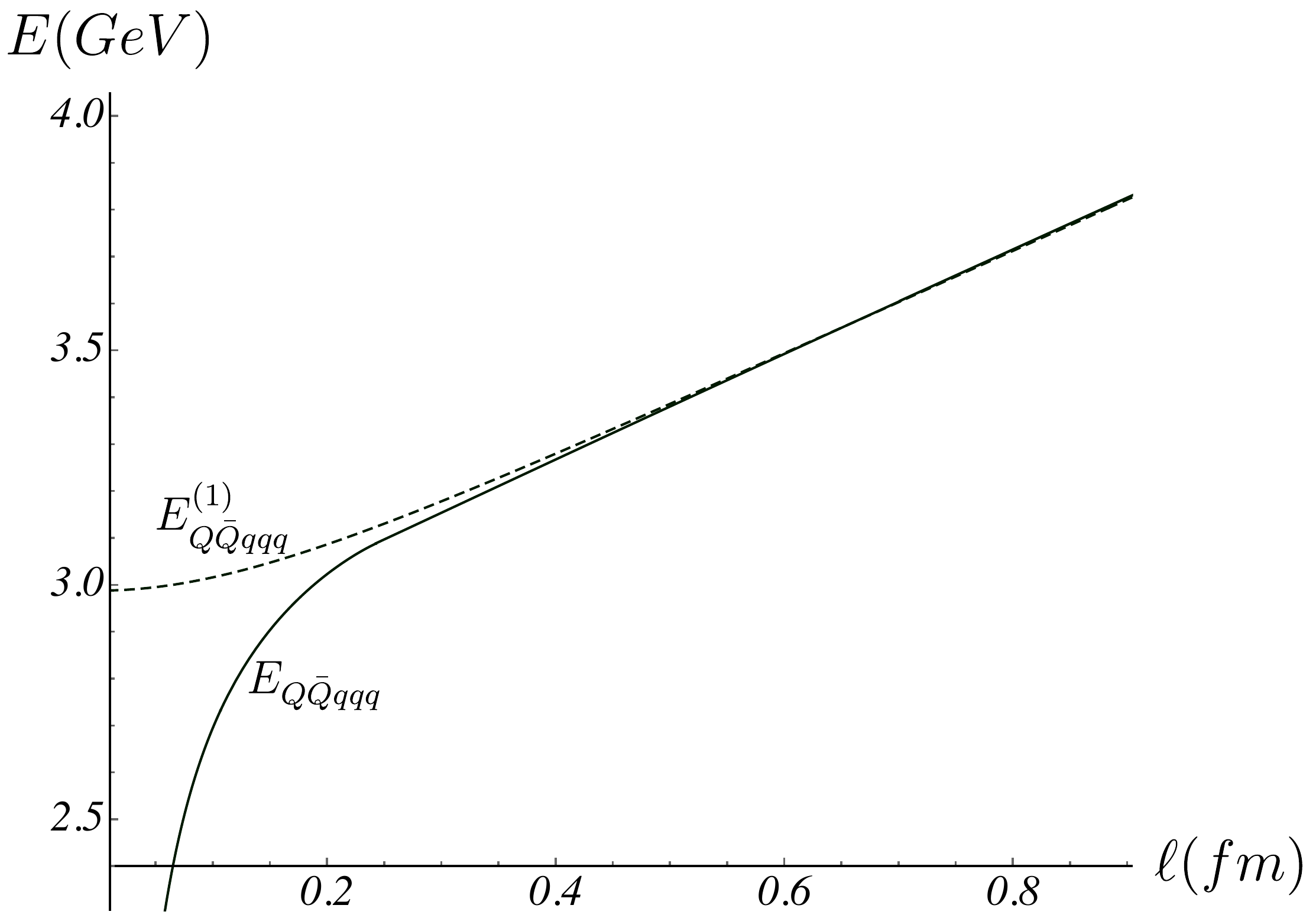}
\hspace{1.3cm}
\includegraphics[width=8.05cm]{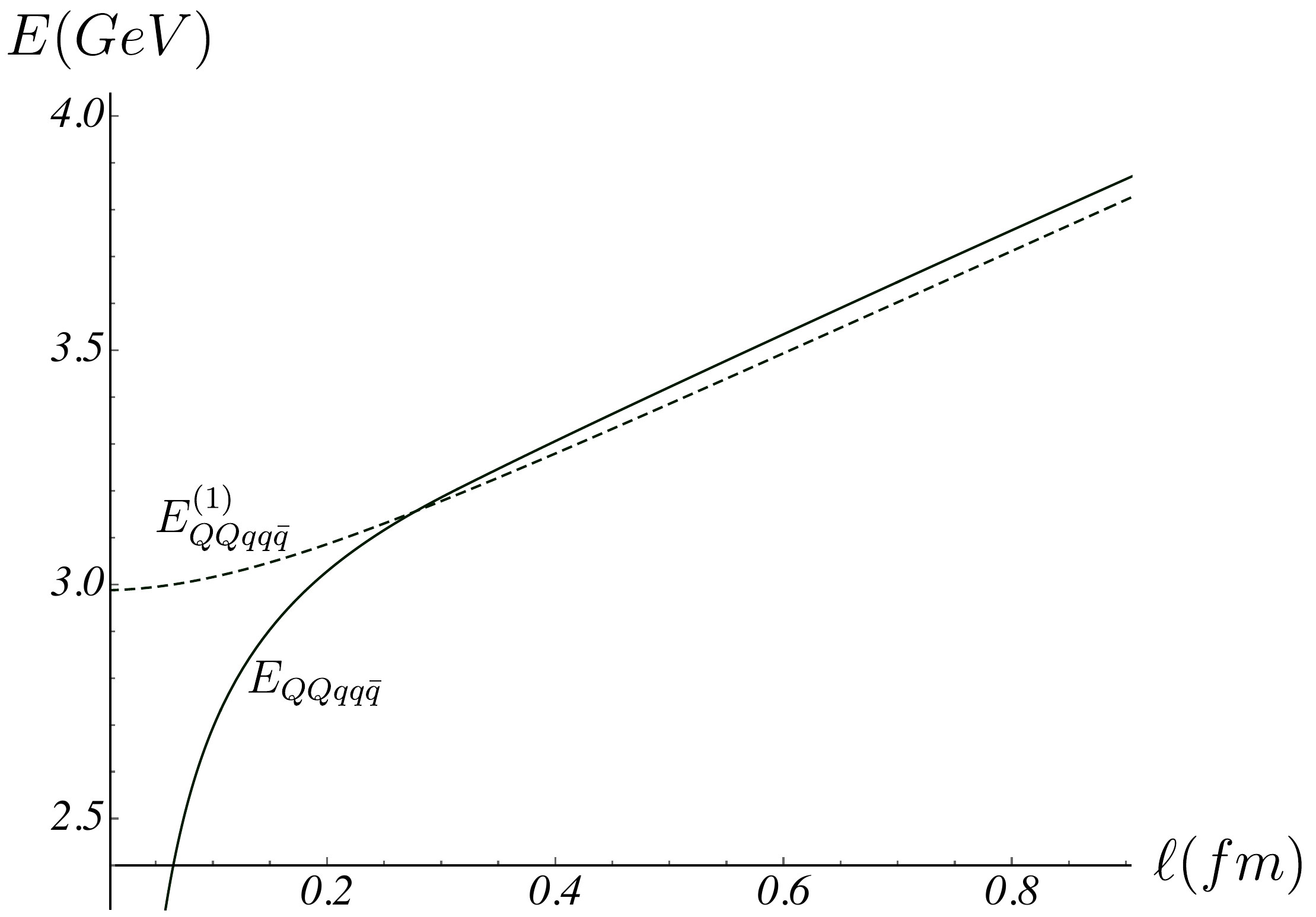}
\caption{{\small Energies of the pentaquark configurations for the $Q\bar Qqqq$ (left) and $QQqq\bar q$ (right) systems.}}
\label{hedgehods}
\end{figure}

It is interesting to ask when the generalized pentaquark configuration may be energetically favorable  (i.e., has lower energy) compared to the standard pentaquark configuration of Figure \ref{c41}. We are now in a position to answer this question. Firstly, it's worth noting that the values of the $\ws$'s can be easily found numerically. As a result we have $\ws\approx 0.625$, $\wsz\approx 0.953$ and $\wso\approx 0.966$, based on the parameter values outlined in Sec. III. Next, in the left panel of Figure \ref{hedgehods}, we plot $E_{\QQbqqq}$ and $E_{\QQbqqq}^{(1)}$ versus $\ell$. As seen from this Figure, the standard pentaquark configuration is

 \noindent favorable at small separations. However, for separations greater than approximately $0.5\,\text{fm}$ the difference between the plots becomes negligible. To determine which configuration has a lower energy, we examine the large-$\ell$ behavior of $E_{\QQbqqb}$ and $E^{(1)}_{\QQbqqq}$. Using the formulas \eqref{EQQbqqq-large} and \eqref{Eh-large}, we find
 
 \begin{equation}\label{Delta-h}
 \Delta=E_{\QQbqqb}^{(1)}-E_{\QQbqqq}=2\g\sqrt{\s}\bigl(I_{\QQbqqq}-I^{(1)}_{\QQbqqq}\bigr)-E_0\,.	
 \end{equation}
A simple estimate yields $\Delta\approx -5\,\text{MeV}$. This implies that the generalized pentaquark configuration is energetically favorable at large separations. Further analysis reveals that the transition between the string configurations occurs at approximately $0.679\,\text{fm}$. It can be interpreted as string junction fusion, as sketched in Figure \ref{sint}(d). 

It is easy to perform the same analysis for the $QQqq\bar q$ system. In the right panel of Figure \ref{hedgehods}, we present the plot of $E_{\QQqqqb}$ and $E^{(1)}_{\QQqqqb}$. Here $E_{\QQqqqb}$ represents the energy of the standard pentaquark configuration (see Figure 2 in \cite{a-QQ3q}). The resulting picture exhibits qualitative similarities to what we got for the $Q\bar Qqqq$ system. However, two quantitative differences are noticeable. First, the gap now is now about $-47\,\text{MeV}$, and thus visible. Second, the transition occurs at a smaller separation, around $0.278\,\text{fm}$.

To summarize, at large separations the generalized pentaquark configurations described by the vertex $V^{(1)}$ have lower energy than the standard ones, but at small separations, the standard ones prevail. The transition between these configurations occurs due to shrinking of strings that eventually leads to string junction fusion: $V\bar VV\rightarrow V^{(1)}$. From the viewpoint of string theory, this is an instance of brane fusion, when several branes coalesce into one (see, e.g., \cite{Dfusion}).\footnote{It is worth noting that junction annihilation can be interpreted as the process of brane annihilation.} 

\subsection{Beyond the Born-Oppenheimer approximation}

In the real world, charm quarks are heavy, but not excessively so, which necessitates the inclusion of finite-mass corrections. If these corrections are substantial, the B-O approximation becomes invalid, and one has to treat light and heavy quarks on an equal footing. If so, then the $c\bar cqqq$ system can be thought of as a subsystem of the $3q$ system by adding a $c\bar c$ pair, which can be considered an excitation. In the case of $uud$, the latter scenario presents a longstanding issue concerning the existence of intrinsic charm quarks within the proton \cite{brodsky-cc}.\footnote{This matter remains a subject of debate in the literature. See, for example, \cite{rev-cc} and references therein.} 

In our discussion, we will focus solely on one aspect: the effective quarkonium-nucleon interaction. According to a proposal in \cite{brodsky}, this interaction is a van der Waals type associated with multiple gluon exchange. The corresponding (non-relativistic) potential is expected to be of the Yukawa form 

\begin{equation}
	V_{(\QQb)\text{A}}=-\frac{\alpha}{L}\ep^{-\mu L}
	\,.
\end{equation}

\begin{figure}[htbp]
\centering
\includegraphics[width=12cm]{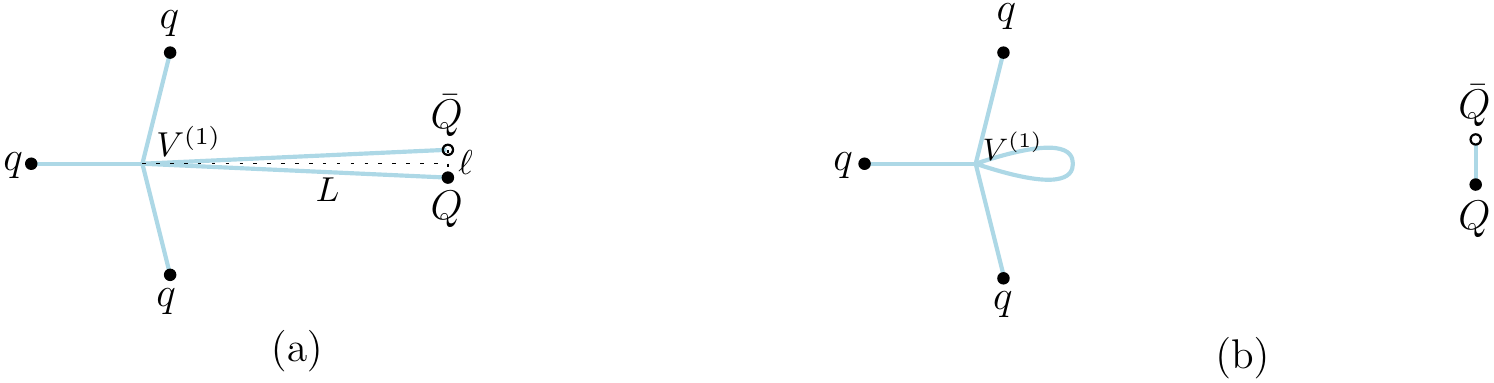}
\caption{{\small Sketched here are string configurations describing the quarkonium-nucleon interaction.}}
\label{QQA}
\end{figure}

We are now in a position to propose a string interpretation of this effective potential.\footnote{Since it is also applicable to a $b\bar b$ pair, we use a general notation $Q$ to denote the $c$ and $b$ quarks.} First of all, it is clear that such a potential is meaningful only if a two-cluster decomposition takes place. In this case, we can define a separation distance $L$ between a nucleon and a $Q\bar Q$ pair as the distance between a baryon vertex and the center of mass of the pair. For small $L$, but enough larger than the separation distance $\ell$ of the $Q\bar Q$ pair, the corresponding string configuration is presented in Figure \ref{QQA}(a).\footnote{It is a special case of the configuration shown in Figure \ref{4h-penta}(a).} The contribution to the potential (binding energy) arises from two \hspace{2cm} 
\vspace{.85 cm}

\noindent strings stretched between the vertex and heavy quarks. Each string gives rise to an attractive Coulomb term.\footnote{To see this in the five-dimensional framework, it is necessary to consider a string stretched between a heavy quark located on the boundary and the vertex in the interior. The Coulomb term can be obtained from the formulas provided in Appendix B of \cite{a-3qPRD}, under the assumption that the tangent angle is not a right angle.} This is the desired result for the effective potential at small separation distances. An important point to note in this argument is that the pair is not a color singlet since the string configuration is connected. The question then arises: what happens for larger values of $L$? Naively, the string contribution leads to a linear potential, but this is not the full story due to string reconnection. In fact, at a fixed value of $\ell$, the separation  between the two oppositely oriented strings decreases as $L$ increases. At some large value of $L$ string reconnection occurs. As a result, the string configuration becomes disconnected, as shown in Figure \ref{QQA}(b). Hence the potential flattens that is consistent with the expected weakening of the quarkonium-nucleon interaction. 

The potential $V_{(\QQb)\text{A}}$ interpolates between the energies of the two string configurations. Formally, it can be defined as the smallest eigenvalue of a model Hamiltonian. The Hamiltonian is represented by a $2\times 2$ matrix

\begin{equation}\label{HQQA}
{\cal H}(L)=
\begin{pmatrix}
E_{\QQbqqq}^{(1)}(L) & \Theta_{\text{A}} \\
\Theta_{\text{A}} & E_{\QQb}+E^{(1)}_{\nucl} \\
\end{pmatrix}
\,,
\end{equation}
where the diagonal elements correspond to the energies of the configurations, and the off-diagonal element describes the mixing between those. 

We conclude this discussion with some remarks. (1) Our scenario is based on the generalized baryon vertex $V^{(1)}$. This assumes that the nucleon is in an excited state, as can be seen from Figure \ref{QQA}(b). (2) For separations of order $\ell$ or smaller, the dominant string configuration is depicted in Figure \ref{c41}, which violates cluster decomposition. (3) Since strings are gluonic objects, what we have just discussed pertains to the pure gluonic contributions to the effective potential. (4) The interaction between the hadrons of Figure \ref{QQA}(b) is encoded in $\Theta_{\text{A}}$. (5) We have only provided a brief outline of the string interpretation of the quarkonium-nucleon interaction, leaving the details for future work.

\section{Further Comments}
\renewcommand{\theequation}{5.\arabic{equation}}
\setcounter{equation}{0}

\subsection{An issue with $E_{\qqb}$ and $E_{\nucl}$}

Drawing conclusions from the plots of Figure \ref{all-L} requires a caveat. As already noted in \cite{a-QQbqqb,a-QQ3q}, the rest energies of the pion and nucleon calculated from the expressions \eqref{Eqqb} and \eqref{nucl} are  $E_{\qqb}=1.190\,\text{GeV}$ and $E_{\nucl}=1.769\,\text{GeV}$. These values differ notably from  the values of $280\,\text{MeV}$ and $1.060\,\text{GeV}$ used in the lattice calculations \cite{bulava}.\footnote{Note that in the case of two flavors $E_{\nucl}=1.060\,\text{GeV}$ at $E_{\qqb}=285\,\text{MeV}$ \cite{nucl}.} The issue is that the effective string model in its current form still does not accurately describe light hadrons because it was originally developed for applications in the heavy quark (static) limit. In the context of string theory on AdS-like geometries, this implies that at least one quark needs to be infinitely massive and positioned on the boundary of five-dimensional space. 

\begin{figure}[H]
\centering
\includegraphics[width=8.75cm]{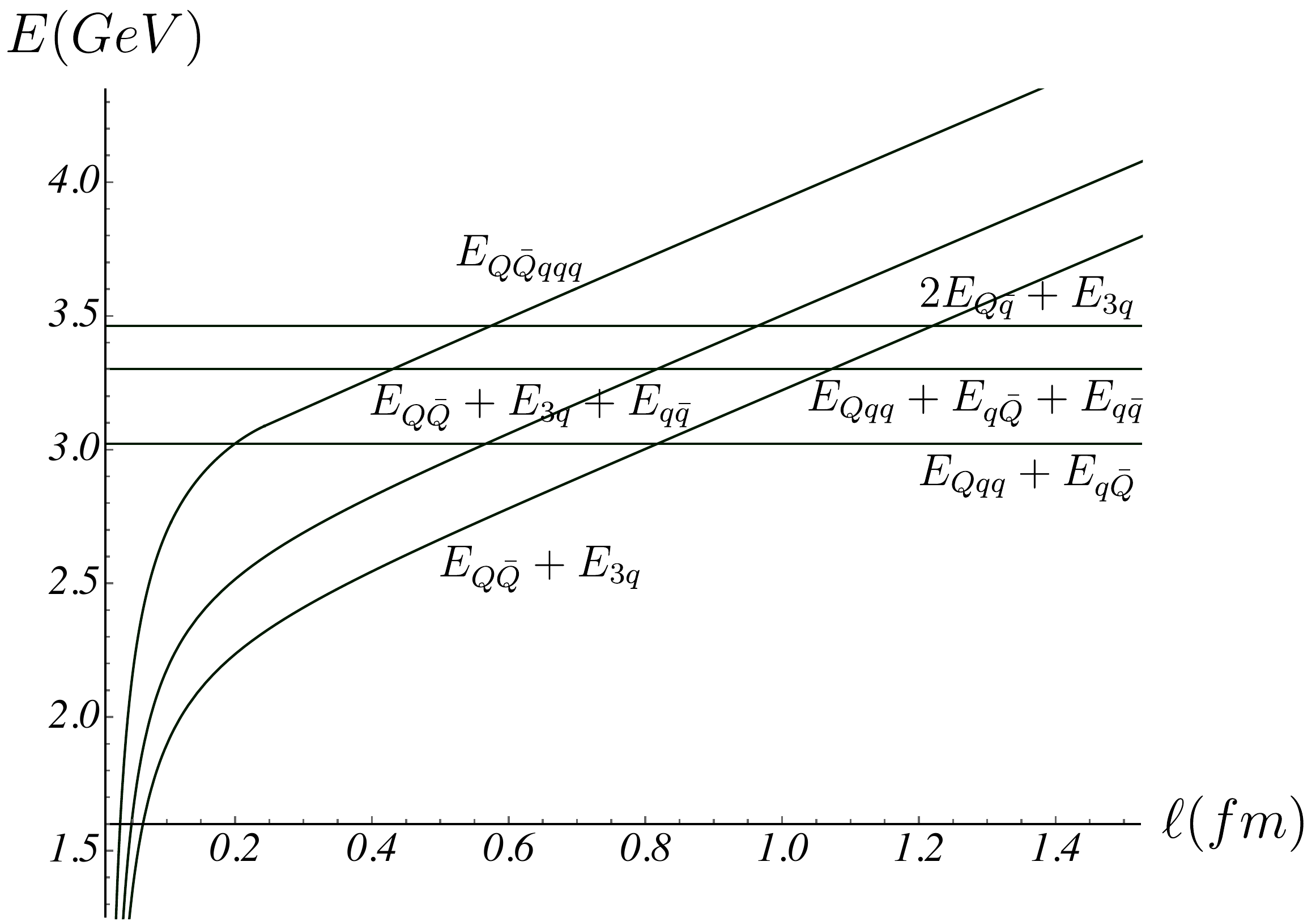}
\hspace{0.3cm}
\includegraphics[width=8.6cm]{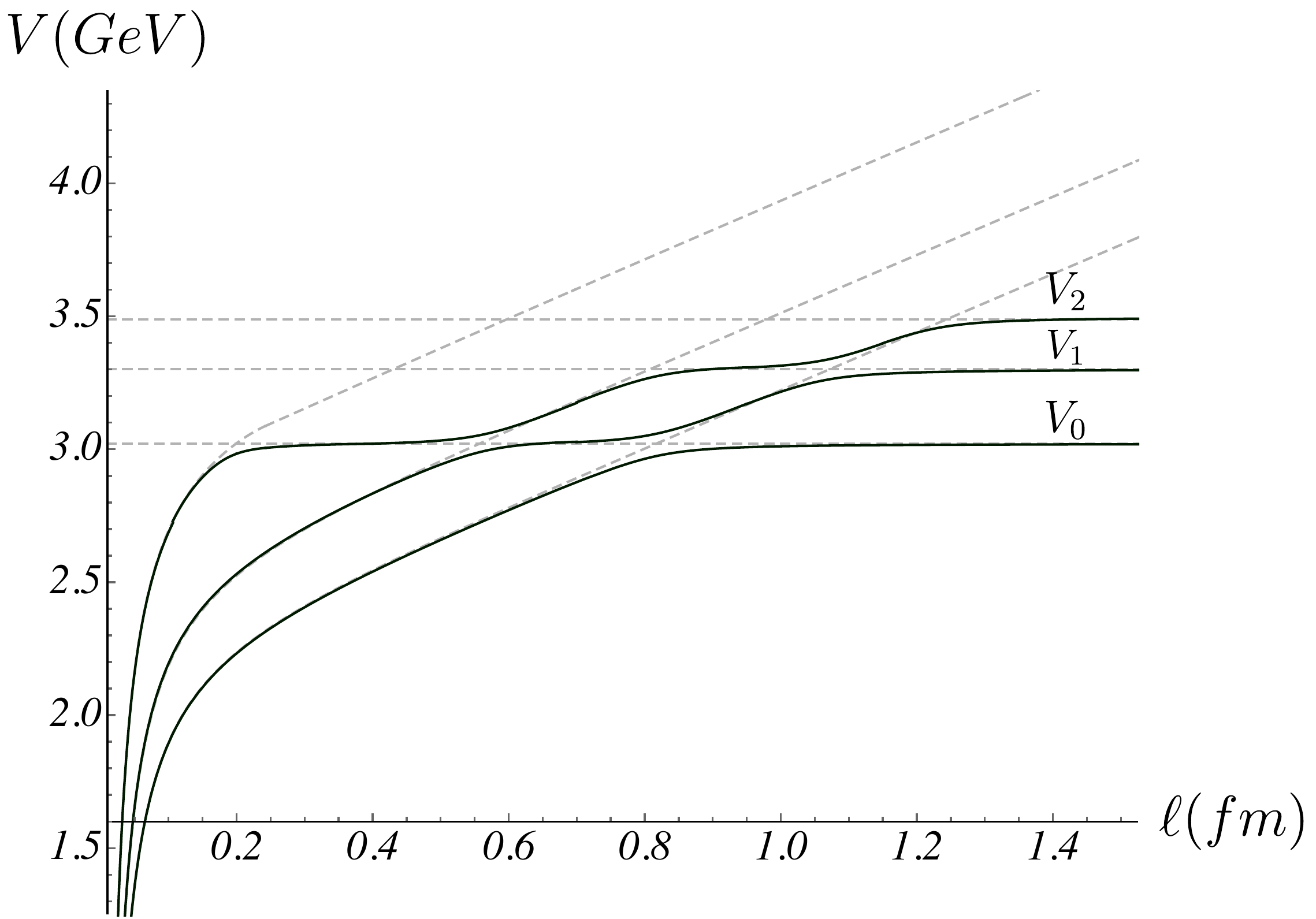}
\caption{{\small Left: The $E$'s vs $\ell$. Right: Sketched here are the three low-lying B-O potentials of the $Q\bar Qqqq$ system. The dashed lines indicate the $E$'s.}} 
\label{Es-pion}
\end{figure}

To some extent, this issue can be addressed by thinking of $E_{\qqb}$ and $E_{\nucl}$ as model parameters \cite{a-QQbqqb,a-QQ3q}. For $E_{\qqb}=280\,\text{MeV}$ and $E_{\nucl}=1.060\,\text{GeV}$, the corresponding $E$'s are plotted in Figure \ref{Es-pion} on the left. The main conclusions that  
\noindent can be drawn from this result are as follows: Configurations (a) and (b) remain the configurations with the lowest energies. The pentaquark configuration (c) now has a higher energy than configuration (d), leading to an interchange of the corresponding graphs compared to Figure \ref{all-L}. A similar interchange also occurs between the graphs for configurations (e) and (f). 

It is worth noting that at almost physical pion mass one argument, based on phenomenological models \cite{MK}, can be made for the validity of $E^{\text{(c)}}>E^{\text{(d)}}$. Adding a pion to configuration (a) results in an energy cost of $145\,\text{MeV}$, whereas adding only one string junction leads to an energy cost of $165\,\text{MeV}$.

For the lowest B-O potential $V_0$, the most visible effect of the change in $E_{\qqb}$ is that string reconnection occurs at a much larger $\ell$, about $0.8\,\text{fm}$.  In this case, Eq.\eqref{lQq} can be approximately solved by using the large-$\ell$ expansion for $E_{\QQb}$. So, with the help of \eqref{EQQb-large}, we find  

\begin{equation}\label{lQq-large}
l_{\Qq}
\approx
\frac{1}{\sigma}
\Bigl(E_{\Qqq}+E_{\Qqb}-2c-E_{\nucl}+2\g\sqrt{\s}I_0\Bigr)
\,.
\end{equation}
In fact, the term $2c$ cancels out due to the $c$-dependent contributions coming from $E_{\Qqq}$ and $E_{\Qqb}$. A numerical estimate shows that $l_{\Qq}\approx0.816\,\text{fm}$. 

The next B-O potential is now defined by $V_1=\min\{E_{\QQb}+E_{\nucl}+E_{\qqb},E_{\Qqq}+E_{\qQb}, E_{\QQb}+E_{\nucl}, E_{\Qqq}+E_{\qQb}+E_{\qqb}\}$. The essential feature of $V_1$ is the emergence of three length scales which separate different configurations. The first scale refers to the process: $Q\bar Q+3q+q\bar q\rightarrow Qqq+q\bar Q$.  In fact, it consists of two subprocesses: virtual pair annihilation and string reconnection. We defines a critical separation distance by 

\begin{equation}\label{lQq-}
E_{\QQb}(l_{\Qq}^{-})+E_{\nucl}+E_{\qqb}=E_{\Qqq}+E_{\Qqb}
\,.
\end{equation}
In $l_{\Qq}^{-}$, the upper subscript refers to virtual pair annihilation. As seen from the Figure, $l_{\Qq}^{-}$ is of order $0.6\,\text{fm}$. Within this range of $\ell$ values, the function $E_{\QQb}(\ell)$ can be approximated by \eqref{EQQb-large}. If so, then a simple calculation shows that 

\begin{equation}\label{lQq-1}
l_{\Qq}^{-}
\approx
\frac{1}{\sigma}
\Bigl(E_{\Qqq}+E_{\Qqb}-2c-E_{\nucl}-E_{\qqb}+2\g\sqrt{\s}I_0\Bigr)
\,.
\end{equation}
It is interesting to estimate the value of $l_{\Qq}^{-}$. For our parameter values, this gives $l_{\Qq}^{-}\approx 0.560\,\text{fm}$. The second scale is related to the process of string reconnection: $Qqq+q\bar Q\rightarrow Q\bar Q+3q$. This process is the inverse of the process of string reconnection discussed above for $V_0$. Because of this, the formula \eqref{lQq-large} holds true. Finally, the third scale is due to the process: $Q\bar Q+3q\rightarrow Qqq+q\bar Q+q\bar q$. It also consists of two subprocesses: virtual pair creation and string reconnection. In this case, we define a critical distance by 

\begin{equation}\label{lQq+}
E_{\QQb}(l_{\Qq}^{+})+E_{\nucl}=E_{\Qqq}+E_{\Qqb}+E_{\qqb}
\,,
\end{equation}
where the upper subscript refers to virtual pair creation. Using the asymptotic formula for $E_{\QQb}$ again, we obtain 

\begin{equation}\label{LQq+1}
l_{\Qq}^{+}
\approx
\frac{1}{\sigma}
\Bigl(E_{\Qqq}+E_{\Qqb}+E_{\qqb}-2c-E_{\nucl}+2\g\sqrt{\s}I_0\Bigr)
\,.
\end{equation}
A simple estimate yields $l_{\Qq}^{+}\approx 1.073\,\text{fm}$. 

As seen from the above Figure, the potential $V_2$ is described in terms of the energies of six different configurations. Therefore, we have $V_2=\min\{E_{\QQbqqq},E_{\Qqq}+E_{\qQb},E_{\QQb}+E_{\nucl}+E_{\qqb}, E_{\Qqq}+E_{\qQb}+E_{\qqb}, E_{\QQb}+E_{\nucl}, 2E_{\Qqb}+E_{\nucl}\}$. There are five emerging scales that we have already analyzed, so we can discuss them briefly. The first transition near $\ell=0.184\,\text{fm}$ is due to string junction annihilation, as discussed in Sec.III. The corresponding critical distance is given by \eqref{sja2}. The second scale refers to the process $Qqq+q\bar Q\rightarrow Q\bar Q+3q+q\bar q$ which is the inverse of the process we have discussed in the case of $V_1$. Because of this, the formula \eqref{lQq-1} is valid. The next transition near $\ell=0.816\,\text{fm}$ is due to string reconnection, with the critical separation distance given by the expression \eqref{lQq-large}. The fourth scale refers to the process $Qqq+q\bar Q+q\bar q\rightarrow Q\bar Q+3q$. It is the inverse of what we have discussed for $V_1$, so the formula \eqref{LQq+1} is applicable for estimating the critical separation distance. Finally, the fifth scale is set by string breaking in the $Q\bar Q$ system, with the corresponding formula in Appendix B.

Two important conclusions that can be drawn from this analysis are: 1) The standard pentaquark configuration provides a dominant contribution to the second excited B-O potential for separations less than $0.2\,\text{fm}$. In that sense, if pentaquarks exist for $V_2$, they are compact. 2) The generalized pentaquark configuration does not contribute to the three low-lying B-O potentials. It becomes relevant for higher excited potentials.

\subsection{More on the potentials}

By understanding the string configurations, we can gain further insight into the three low-lying B-O potentials. In doing so, we follow the approach of lattice QCD, which is commonly used to investigate string breaking in the $Q\bar Q$ system \cite{FK}, and consider a model Hamiltonian which for the problem at hand is a $6\times 6$ matrix. Explicitly, 

\begin{equation}\label{HV012}
{\cal H}(\ell)=
\begin{pmatrix}
\,E_{\QQb}(\ell)+E_{\nucl} & {} & {} & {} & {} & {}\, \\
\,{} & E_{\Qqq}+E_{\qQb} & {} & {} & {} & {} \,\\
\,{}& {} & E_{\QQb}(\ell)+E_{\nucl}+E_{\qqb} & {} & {} & \Theta_{ij}\, \\
\,{} & {} & {} & E_{\Qqb}+E_{\Qqq}+E_{\qqb} & {} & {}\,\\
\,\Theta_{ij} & {} & {} & {} & E_{\QQbqqq}(\ell) & {} \,\\
\,{} & {} & {} & {} & {} & 2E_{\Qqb}+E_{\nucl}\,
\end{pmatrix}
\,,
\end{equation}
where the off-diagonal elements describe the strength of mixing between the six distinct states (string configurations). The first three low-lying B-O potentials correspond to the three smallest eigenvalues of the matrix ${\cal H}$. 

Unlike lattice QCD, where the Hamiltonian can potentially be determined from a correlation matrix, it remains unclear how to calculate the off-diagonal elements within the effective string model. Consequently, it becomes challenging to precisely visualize the form of the potentials. However, we can gain insight from our previous experiences with other quark systems regarding the approximate magnitudes of the $\Theta$ values near the transition points.\footnote{For instance, we can assume these $\Theta$ values to be approximately $47\,\text{MeV}$, as in the $Q\bar Q$ system on the lattice \cite{bulava}.} By doing so, the overall picture becomes more akin to the one sketched in Figure \ref{Es-pion} on the right. The compact pentaquark configuration predominantly contributes to the potential $V_2$ for heavy quark separations smaller than $0.2\,\text{fm}$.

\subsection{A relation among hadron masses}

Using Eqs.\eqref{Qqb} and \eqref{EQqq}, we can rewrite the relation \eqref{maiani} as follows:

\begin{equation}\label{PT}
E_{\QQbqqq}(\ell)=E_{\QQqbqb}(\ell)+E_{\Qqq}-E_{\Qqb}  
\,.
\end{equation}
We assume also that $E_{\QbQbqq}=E_{\QQqbqb}$. It is tempting to apply this relation in the heavy quark limit, where contributions from the motion of the heavy quarks and spin interactions are negligible, to derive a relation among the masses of doubly-heavy-light and heavy-light hadrons. This gives

\begin{equation}\label{MPT}
m_{\QQbqqq}-m_{\QQqbqb}=m_{\Qqq}-m_{\Qqb}
\,.
\end{equation}
It is necessary to keep in mind that because $E_{\QQbqqq}$ and $E_{\QQqbqb}$ provide the dominant contributions to the potentials at small heavy quark separations, the doubly heavy hadrons are compact in sense of heavy quark separation. Moreover, they are assumed to be described by the connected string configurations.

Interestingly, a similar relation is known for the $QQ\bar q \bar q$ quark system \cite{quigg}, where

\begin{equation}\label{quigg}
m_{\QQqbqb}-m_{\QQq}=m_{\Qqq}-m_{\Qqb}
\,.
\end{equation}
It can be derived from heavy quark-diquark symmetry \cite{wise} that also assumes that the doubly heavy hadrons are compact. 

It is intriguing to examine the possible phenomenological implications of \eqref{MPT}. Since it is derived in the heavy quark limit, it is natural to attempt some estimates of the masses of hidden-bottom pentaquarks. For brevity, we will not discuss all such pentaquarks here, but only provide a simple estimate for the lightest one. In this case, several predictions can be found in the literature \cite{wu,yang,ferr,valc}. We compare those with our estimate based on \eqref{MPT}. The result is presented in Table \ref{estimates}. 
\begin{table*}[htb]
\renewcommand{\arraystretch}{2}
\centering 	
\begin{tabular}{lccccccr}				
\hline
\hline
State ~~~~~&~~~~~~\cite{ferr} ~~~~~&~~~Our model~~&~~~~~\cite{valc}~~~~~& ~~~~~\cite{yang}~~~~~&~~~\cite{wu}~~
\rule[-3mm]{0mm}{8mm}
\\
\hline 
$b\bar bqqq$ & 10.605 & 10.821 & 11.062 & 11.080 & 11.137 \\
\hline
\hline
\end{tabular}
\caption{ \small Predictions of different models for the mass (in GeV) of the lightest pentaquark state.}
\label{estimates}
\end{table*}
In the process we used the hadron masses from \cite{quigg} which include $m_{bb\bar q\bar q}=10.482\,\text{GeV}$, $m_{bqq}=5.619\,\text{GeV}$, and $m_{b\bar q}=5.28\,\text{GeV}$. Although the obtained value falls within the range of the recent predictions, only the experiment can definitively ascertain the mass.

\section{Conclusions and Outlook}
\renewcommand{\theequation}{6.\arabic{equation}}
\setcounter{equation}{0}

The somewhat surprising conclusions regarding the $QQqq\bar q$ and $Q\bar Qqqq$ pentaquark systems provide strong evidence that the ground state B-O potential is described in terms of hadro-quarkonia and hadronic molecules. The transition between the hadro-quarkonium description at small separation distances and molecular description at larger separations occurs due to the phenomenon of string reconnection at a critical separation distance of approximately $0.8-1.0\,\text{fm}$. Pentaquark states described by genuine five-quark interactions may appear in the case of higher lying B-O potentials, namely $V_1$ or $V_2$, depending on the specific  model being used. These states are compact in the sense of small separations between heavy quark sources. This should be a useful guide when analyzing the nature of the doubly heavy pentaquark systems.  

There are still several important problems that need to be addressed in order to establish contact with the real world. Among these problems are:

(i) The treatment of the off-diagonal elements $\Theta$ in the model Hamiltonians as model parameters. It is highly desirable to develop a string theory technique that enables a direct computation of these elements.

(ii) Due to the lack of lattice simulations, we have relied on available data at nearly the double value of the pion mass to fix the model parameters. This issue requires further attention. The recent work \cite{aoki} gives just one example of needed improvements.  The length scale of string junction annihilation is noticeably larger at $m_{\pi}=146\,\text{MeV}$.

(iii) It is important to go beyond the heavy quark limit and estimate the leading $1/m_Q$ corrections, especially in the case of the $c$ quark.  

(iv) The five-dimensional string model considered in this paper is inspired by string theory in ten dimensions. In the context of the latter, the string junction is a five-brane. This five-brane can exist not only in its ground state but also in excited states. Furthermore, it allows strings to both begin and end on the brane, leading to the concept of generalized string junctions, as discussed in Section IV. The main question is whether any evidence of the brane nature of the string junction can be observed in lattice QCD. In addition to investigating the hybrid potentials of the $3Q$ system, another avenue to explore such evidence is by examining the pentaquark system through the study of the Wilson loop defined as\footnote{At this point, it's noteworthy that assuming the analogy between Wilson lines and strings, one can immediately infer that, apart from the standard string junction $V^{(0)}$, a new junction $V^{(1)}$ would arise, where five strings meet. A similar line of reasoning can be applied to a multi-string junction $V^{(N)}$.}  
\begin{equation}\label{W1}
W^{(1)}_{5Q}=P^{abcd}_kP^{k'}_{a'b'c'd'} U_a^{a'}U_b^{b'}U_c^{c'}U_d^{d'}U_{k'}^k
\,.
\end{equation}
Here the $U$'s represent the path-ordered exponents similar to those in a baryonic Wilson loop, but with the last exponent oriented in the opposite direction to the first four. The tensor $P$ is a combination of the $\varepsilon$ and $\delta$ tensors. For example, $P^{abcd}_k=\delta^a_k\varepsilon^{bcd}$ or $P^{abcd}_k=\delta^a_k\varepsilon^{bcd}-\delta_k^b\varepsilon^{acd}$.

\begin{acknowledgments}
We would like to thank S. Aoki, S. Krippendorf, J. Sonnenschein, and P. Weisz for useful discussions. This work was supported in part by Russian Science Foundation grant 20-12-00200 in association with Steklov Mathematical Institute.

\end{acknowledgments}
\appendix
\section{Notation}
\renewcommand{\theequation}{A.\arabic{equation}}
\setcounter{equation}{0}
Throughout the paper, heavy and light quarks (antiquarks) are denoted by $Q(\bar Q)$ and $q(\bar q)$ respectively, and baryon (antibaryon) vertices by $V(\bar V)$. Light quarks (antiquarks) are located at $r=\rq(\rqb)$, while vertices at $r=\rv(\rvb)$ unless otherwise specified. It is convenient to introduce dimensionless variables: $q=\s\rq^2$, $\bar q=\s\rqb^2$, $v=\s\rv^2$, and $\bar v=\s\rvb^2$. These variables range from 0 to 1 and indicate the proximity of the objects to the soft-wall, which is located at 1 in such units. To classify the critical separations related to the string interactions depicted in Figure \ref{sint}, we use the notation $l$ for (a), $\ell$ for (b), and $\boldsymbol{\ell}$ for (c).

In order to express the resulting formulas concisely, we utilize the set of basic functions \cite{a-stb3q}:

\begin{equation}\label{fL+}
{\cal L}^+(\alpha,x)=\cos\alpha\sqrt{x}\int^1_0 du\, u^2\, \ep^{x (1-u^2)}
\Bigl[1-\cos^2{}\hspace{-1mm}\alpha\, u^4\ep^{2x(1-u^2)}\Bigr]^{-\frac{1}{2}}
\,,
\qquad
0\leq\alpha\leq\frac{\pi}{2}\,,
\qquad 
0\leq x\leq 1
\,.
\end{equation}
${\cal L}^+$ is a non-negative function which vanishes if $\alpha=\frac{\pi}{2}$ or $x=0$, and has a singular point at $(0,1)$. Assuming that $\alpha$ is a function of $x$ such that $\cos\alpha(x)=\cos\alpha+\cos'\hspace{-0.9mm}\alpha x+o(x)$ as $x\rightarrow 0$, the small-$x$ behavior of ${\cal L}^+$ is 

\begin{equation}\label{fL+smallx}
{\cal L}^+(\alpha,x)=\sqrt{x}
\bigl({\cal L}^+_0+{\cal L}^+_1 x+o(x)\bigr)
\,,
\end{equation}
where 
\begin{equation*}
{\cal L}^+_0=\frac{1}{4}
\cos^{-\oh}\hspace{-.9mm}\alpha\,B\bigl(\cos^2\hspace{-.9mm}\alpha;\tfrac{3}{4},\tfrac{1}{2}\bigr)
\,,\qquad
{\cal L}^+_1=\frac{1}{4}
\cos^{-\frac{3}{2}}\hspace{-.9mm}\alpha
\Bigl(
(\cos\alpha+\cos'\hspace{-.9mm}\alpha\bigr)
B\bigl(\cos^2\hspace{-.9mm}\alpha;\tfrac{3}{4},-\tfrac{1}{2}\bigr)-B\bigl(\cos^2\hspace{-.9mm}\alpha;\tfrac{5}{4},-\tfrac{1}{2}\bigr)
\Bigr)
\,,
\end{equation*}
and $B(z;a,b)$ is the incomplete beta function;

\begin{equation}\label{fL-}
{\cal L}^-(y,x)=\sqrt{y}
\biggl(\,
\int^1_0 du\, u^2\, \ep^{y(1-u^2)}
\Bigl[1-u^4\,\ep^{2y(1-u^2)}\Bigr]^{-\frac{1}{2}}
+
\int^1_
{\sqrt{\frac{x}{y}}} 
du\, u^2\, \ep^{y(1-u^2)}
\Bigl[1-u^4\,\ep^{2y(1-u^2)}\Bigr]^{-\frac{1}{2}}
\,\biggr)
\,,
\quad
0\leq x\leq y\leq 1
\,.
\end{equation}
 This function is non-negative and equals zero at the origin, but it becomes singular at $y=1$. Notice that near $y=1$, with $x$ kept fixed, it behaves as

\begin{equation}\label{L-y=1}
{\cal L}^-(y,x)=-\ln(1-y)+O(1)\,.
\end{equation}
The ${\cal L}$ functions are related as ${\cal L}^+(0,x)={\cal L}^-(x,x)$;

\begin{equation}\label{fE+}
{\cal E}^+(\alpha,x)=
\frac{1}{\sqrt{x}}
\int^1_0\,\frac{du}{u^2}\,\biggl(\ep^{x u^2}
\Bigl[
1-\cos^2{}\hspace{-1mm}\alpha\,u^4\ep^{2x (1-u^2)}
\Bigr]^{-\frac{1}{2}}-1-u^2\biggr)
\,,
\qquad
0\leq\alpha\leq\frac{\pi}{2}\,,
\qquad 
0\leq x\leq 1
\,.
\end{equation}
${\cal E}^+$ is singular at $x=0$ and $(0,1)$. If $\cos\alpha(x)=\cos\alpha+\cos'\hspace{-0.9mm}\alpha x+o(x)$ as $x\rightarrow 0$, then the small-$x$ behavior of ${\cal E}^+$ is 

\begin{equation}\label{fE+smallx}
{\cal E}^+(\alpha,x)=\frac{1}{\sqrt{x}}\Bigl({\cal E}^+_0+{\cal E}^+_1x+o(x)\Bigr)
\,,
\end{equation}
where
\begin{equation*}
{\cal E}_0^+=
\frac{1}{4}
\cos^{\oh}\hspace{-.9mm}\alpha\,B\bigl(\cos^2\hspace{-.9mm}\alpha;-\tfrac{1}{4},\tfrac{1}{2}\bigr)
\,,\quad
{\cal E}^+_1=\frac{1}{4}
\cos^{-\oh}\hspace{-.9mm}\alpha
\biggl(
(\cos\alpha+\cos'\hspace{-.9mm}\alpha\bigr)B\bigl(\cos^2\hspace{-.9mm}\alpha;\tfrac{3}{4},-\tfrac{1}{2}\bigr)
-3B\bigl(\cos^2\hspace{-.9mm}\alpha;\tfrac{5}{4},-\tfrac{1}{2}\bigr)
+
4\frac{\cos^{\oh}\hspace{-.9mm}\alpha}{\sin\alpha}
\biggr)
\,;
\end{equation*}

\begin{equation}\label{fE-}
{\cal E}^-(y,x)=\frac{1}{\sqrt{y}}
\biggl(
\int^1_0\,\frac{du}{u^2}\,
\Bigl(\ep^{y u^2}\Bigl[1-u^4\,\ep^{2y(1-u^2)}\Bigr]^{-\frac{1}{2}}
-1-u^2\Bigr)
+
\int^1_{\sqrt{\frac{x}{y}}}\,\frac{du}{u^2}\,\ep^{y u^2}
\Bigl[1-u^4\,\ep^{2y(1-u^2)}\Bigr]^{-\frac{1}{2}}
\biggr) 
\,,
\,\,\,
0\leq x\leq y\leq 1
\,.
\end{equation}
${\cal E}^-$ is singular at $(0,0)$ and at $y=1$. More specifically, near $y=1$, with $x$ kept fixed, it behaves as

\begin{equation}
	{\cal E}^-(y,x)=-\ep\ln(1-y)+O(1)
	\,.
\end{equation}
The ${\cal E}$ functions are also related as ${\cal E}^+(0,x)={\cal E}^-(x,x)$;

\begin{equation}\label{Q}
{\cal Q}(x)=\sqrt{\pi}\text{erfi}(\sqrt{x})-\frac{\ep^x}{\sqrt{x}}
\,.
\end{equation}
Here $\text{erfi}(x)$ is the imaginary error function. This is a special case of ${\cal E}^+$ with $\alpha=\frac{\pi}{2}$.  A useful fact is that its small-$x$ behavior is given by 

\begin{equation}\label{Q0}
{\cal Q}(x)=-\frac{1}{\sqrt{x}}+\sqrt{x}+O(x^{\frac{3}{2}})
\,;
\end{equation}

\begin{equation}\label{I}
	{\cal I}(x)=
	I_0
	-
	\int_{\sqrt{x}}^1\frac{du}{u^2}\ep^{u^2}\Bigl[1-u^4\ep^{2(1-u^2)}\Bigr]^{\frac{1}{2}}
	\,,
\quad\text{with}\quad 
I_0=\int_0^1\frac{du}{u^2}\Bigl(1+u^2-\ep^{u^2}\Bigl[1-u^4\ep^{2(1-u^2)}\Bigr]^{\frac{1}{2}}\Bigr)
\,,
\qquad
0< x\leq 1
\,.
\end{equation}
Notice that $I_0$ can be evaluated numerically, with the result $0.751$.

\section{The potential $V_0$ of the $Q\bar Q$ system}
\renewcommand{\theequation}{B.\arabic{equation}}
\setcounter{equation}{0}

In this Appendix we give a brief summary of the basic results about the ground state B-O potential of a static quark-antiquark pair in the presence of two light flavors of equal mass. These results are pertinent to our discussion in Section III. For standard explanations, see \cite{az1,a-strb} whose conventions we follow, unless otherwise stated.

\begin{figure}[H]
\centering
\includegraphics[width=8.25cm]{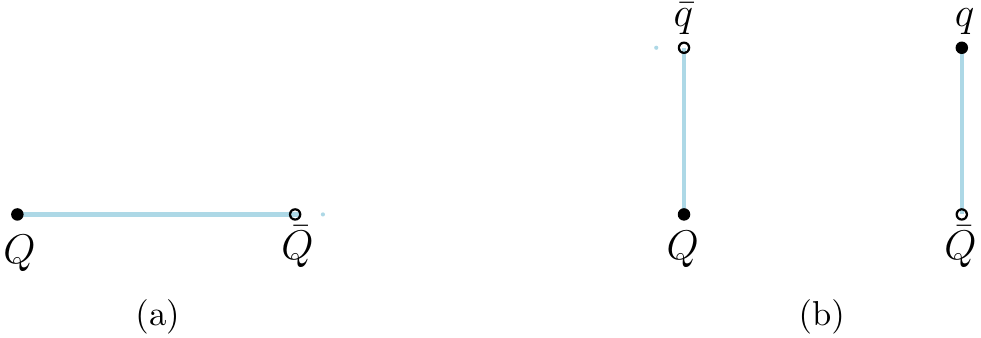}
\caption{{\small The string configurations which contribute to the potential $V_0$ of the $Q\bar Q$ system.}}
\label{4QQb}
\end{figure}

From the perspective of four-dimensional string models \cite{XA}, the only relevant string configurations are those shown in Figure \ref{4QQb}. The first configuration is the simplest connected one, consisting of a valence quark and an antiquark 
 \noindent joined by a string. The second configuration is disconnected, and is formed by adding a pair of light quarks and attaching strings to the quarks in a way that results in a pair of heavy-light mesons. These configurations have a physical meaning: in the context of string models, a heavy meson decay $Q\bar Q\rightarrow Q\bar q+q\bar Q$ is described  as a transition between the two configurations. The transition occurs because of the process of string breaking. The key feature of this process is the creation of a light quark-antiquark pair.

\begin{figure}[htbp]
\centering
\includegraphics[width=11cm]{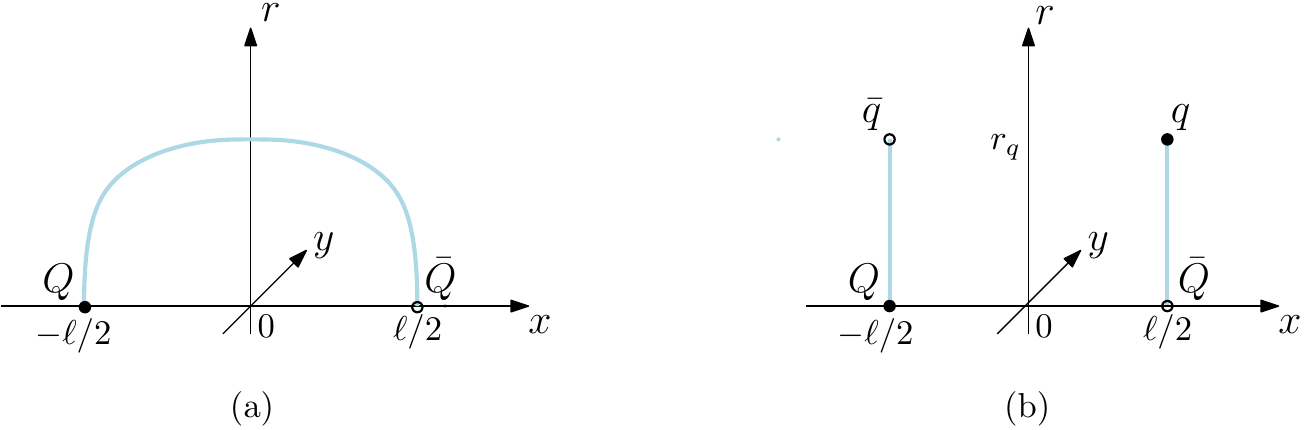}
\caption{{\small The five-dimensional counterparts to the string configurations of Figure \ref{4QQb}.}}
\label{5QQb}
\end{figure}

In five dimensions, the connected configuration consists of a string that is attached to the heavy quark sources located on the boundary of the five-dimensional space, as depicted in Figure \ref{5QQb}(a). 
For the geometry described by Eq.\eqref{metric}, the relation between the quark separation distance along the $x$-axis and the string energy is written in parametric form 

\begin{equation}\label{EQQb}
\ell= \frac{2}{\sqrt{\s}}{\cal L}^+(0,v)
 \,,
\quad
E_{\QQb}=2\g\sqrt{\s}\,{\cal E}^+(0,v)+2c\,.
\end{equation}
Here $v$ is a dimensionless parameter running from $0$ to $1$ and $c$ is the normalization constant as before. The functions ${\cal L}^+$ and ${\cal E}^+$ are defined  in Appendix A.

For future reference, we note that the small-$\ell$ behavior of $E_{\QQb}$ is given by 

\begin{equation}\label{EQQb-small}
E_{\QQb}(\ell)=-\frac{\alpha_{\QQb}}{\ell}+2c+\boldsymbol{\sigma}_{\QQb}\ell +o(\ell)
\,,
\end{equation}
with 
\begin{equation}\label{alpha-QQb}
\alpha_{\QQb}=(2\pi)^3\Gamma^{-4}\bigl(\tfrac{1}{4}\bigr)\g
	\,,\qquad
	\boldsymbol{\sigma}_{\QQb}=\oh(2\pi)^{-2}\Gamma^{4}\bigl(\tfrac{1}{4}\bigr)\g\s
	\,.
\end{equation}
On the other hand, the large-$\ell$ behavior is  

\begin{equation}\label{EQQb-large}
E_{\QQb}(\ell)=\sigma\ell-2\g\sqrt{\s}\,I_0+2c+o(1)
\,,\qquad
\text{with}
\qquad
\sigma=\ep\g\s 
\,.
\end{equation}
Here $I_0$ is defined in Appendix A and $\sigma$ is the physical string tension. It has a larger value than the coefficient $\boldsymbol{\sigma}_{\QQb}$. Numerically, the ratio of $\boldsymbol{\sigma}_{\QQb}$ to $\sigma$ is approximately $0.805$.

Figure \ref{5QQb}(b) represents the five-dimensional counterpart of the disconnected configuration of Figure \ref{4QQb}(b). Since the mesons are non-interacting, the energy is just twice the heavy-light meson mass $E_{\Qqb}$. The latter is given by Eq.\eqref{Qqb}. 

The ground state B-O potential is formally defined by $V_{0}=\min\bigl(E_{\QQb},2E_{\Qqb}\bigr)$. Thus $V_0$ varies between $E_{\QQb}$ at small quark separations and $2E_{\Qqb}$ at larger separations. However, this formal definition does not precisely describe what happens at intermediate quark separations. To address this issue, we can use the same mixing analysis as in lattice gauge theory \cite{FK,bulava}. Specifically, consider a model Hamiltonian of a two-state system 

\begin{equation}\label{HD-QQb}
{\cal H}(\ell)=
\begin{pmatrix}
E_{\QQb}(\ell) & \Theta_{\QQb} \\
\Theta_{\QQb} & 2E_{\Qqb} \\
\end{pmatrix}
\,,
\end{equation}
with $\Theta_{\QQb}$ describing the mixing between the two states. The potential is then obtained as the smallest eigenvalue of the model Hamiltonian. Explicitly, 

\begin{equation}\label{V0QQb}
V_{0}=\oh\Bigl(E_{\QQb}+2E_{\Qqb}\Bigr)
-
\sqrt{\frac{1}{4}\Bigl(E_{\QQb}-2E_{\Qqb}\Bigr)^2+\Theta_{\QQb}^2}
\,.	
\end{equation}

Just like in lattice gauge theory \cite{bulava}, the critical separation distance (often called the string breaking distance) is defined by equating the energies of the configurations

\begin{equation}\label{LcQQb}
E_{\QQb}(\ell_{\QQb})=2E_{\Qqb}
\,.
\end{equation}
This distance provides a condition for determining which configuration is dominant in the system's ground state at given heavy quark separation. For large quark separations, $E_{\QQb}(\ell)$ becomes a linear function of $\ell$ and, as a result, the equation drastically simplifies.\footnote{For the parameter values we use, this is true for $\ell \gtrsim 0.5\,\text{fm}$, whereas the string breaking distance is about $1\,\text{fm}$.} If so, then it follows from Eqs.\eqref{Qqb} and \eqref{EQQb-large} that the string breaking distance is  

\begin{equation}\label{LQQb-large}
\ell_{\QQb}\approx
\frac{2}{\ep\sqrt{\s}}
\Bigl(
{\cal Q}(\qs)+\n\frac{\ep^{\oh \qs}}{\sqrt{\qs}}+I_0
\Bigr)
\,.
\end{equation}
Here $\qs$ is a solution to Eq.\eqref{q}. 

\begin{figure}[H]
\centering
\includegraphics[width=7.85cm]{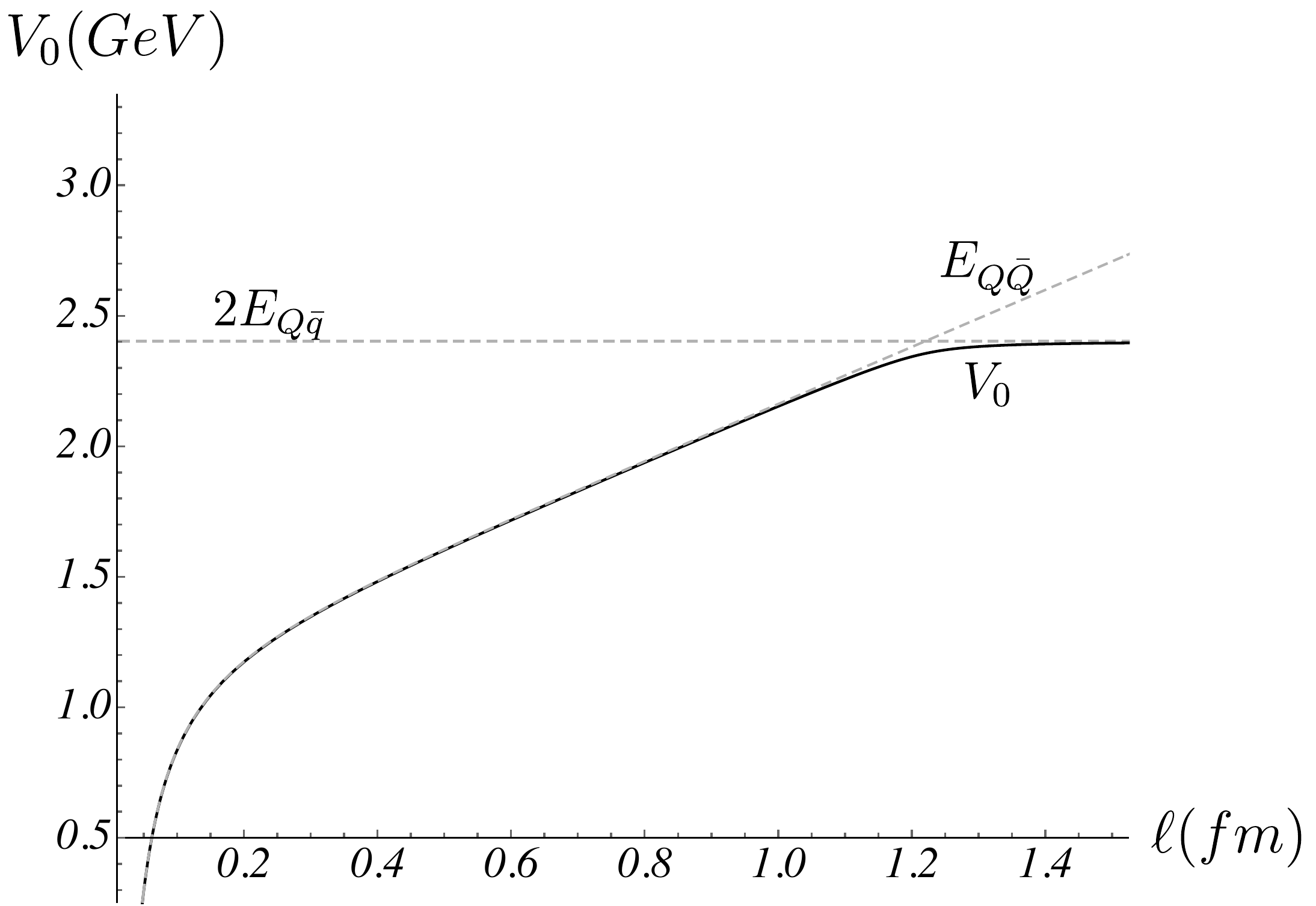}
\caption{{\small The potential determined using Eq.\eqref{V0QQb} with $\Theta_{\QQb}=47\,\text{MeV}$ from \cite{bulava}. The $E$'s are shown in dashed lines.}}
\label{VQQ}
\end{figure}

To illustrate this construction, we will provide a simple example. For the parameter values set in Section III, and with a constant $\Theta_{\QQb}$, the potential is depicted in Figure \ref{VQQ}. We see that as $\ell$ approaches zero, $V_0$ asymptotically app- 
\noindent roaches $E_{\QQb}$, while as $\ell$ tends towards infinity, it approaches $2E_{\Qqb}$. The transition between these two regimes occurs around $\ell=1.22\,\text{fm}$, which is in line with our expectations \cite{bulava}.

\section{Details for the pentaquark configuration of Figure \ref{c52}(c')}
\renewcommand{\theequation}{C.\arabic{equation}}
\setcounter{equation}{0}

To get to the specific issues of interest here as quickly as possible, we will use the fact that the action which governs configuration (c') follows from \eqref{Sc11} at $v=\bar v$. So, we have 

\begin{equation}\label{Sc-w2}
S=3\g T
\biggl(
\frac{2}{3}\int_{0}^{\rvb} \frac{dr}{r^2}\,\ep^{\s r^2}\sqrt{1+(\partial_r x)^2}\,\,
+
\int_{\rvb}^{\rq} \frac{dr}{r^2}\,\ep^{\s r^2}
+
3\k\,\frac{\ep^{-2\s\rvb^2}}{\rvb}
+
\n\frac{\ep^{\frac{1}{2}\s\rq^2}}{\rq}
\,\biggr)
\,. 
\end{equation}
If we vary this action with respect to the position of the light quarks, this will lead us to Eq.\eqref{q}. But if we vary it with respect to the position of the vertices, then we get the equation  

\begin{equation}\label{alpha2}
\sin\alpha=\frac{3}{2}\Bigl(1+3\k(1+4\bar v)\ep^{-3\bar v}\Bigr)
\,,
\end{equation}
which differs from Eq.\eqref{alpha1} by the factor $\frac{3}{2}$. This factor will be crucial for our analysis below. By by the same sort of argument given in subsection B of Sec.III, the expressions for the separation distance and energy are given by Eqs. \eqref{lc-s} and \eqref{Ec-s} with $\vs=\bar v$. The latter now takes the form

\begin{equation}\label{Ec-s12}
E'_{\QQbqqq}=3\g\sqrt{\s}
\biggl(
\frac{2}{3}{\cal E}^+(\alpha,\bar v)
+
{\cal Q}(\qs)-{\cal Q}(\bar v)
+
3\k\frac{\ep^{-2\bar v}}{\sqrt{\bar v}}
+
\n\frac{\ep^{\oh \qs}}{\sqrt{\qs}}
\biggr)
+2c
\,.
\end{equation}
Here we use the prime to highlight the energy of configuration (c') as opposed to that of configuration (c). 

Thus the energy of the configuration is given in parametric form by $E'_{\QQbqqq}=E'_{\QQbqqq}(\bar v)$ and $\ell=\ell(\bar v)$, with the parameter varying from $\vs$ to $\qs$. 

\begin{figure}[htbp]
\centering
\includegraphics[width=5.8cm]{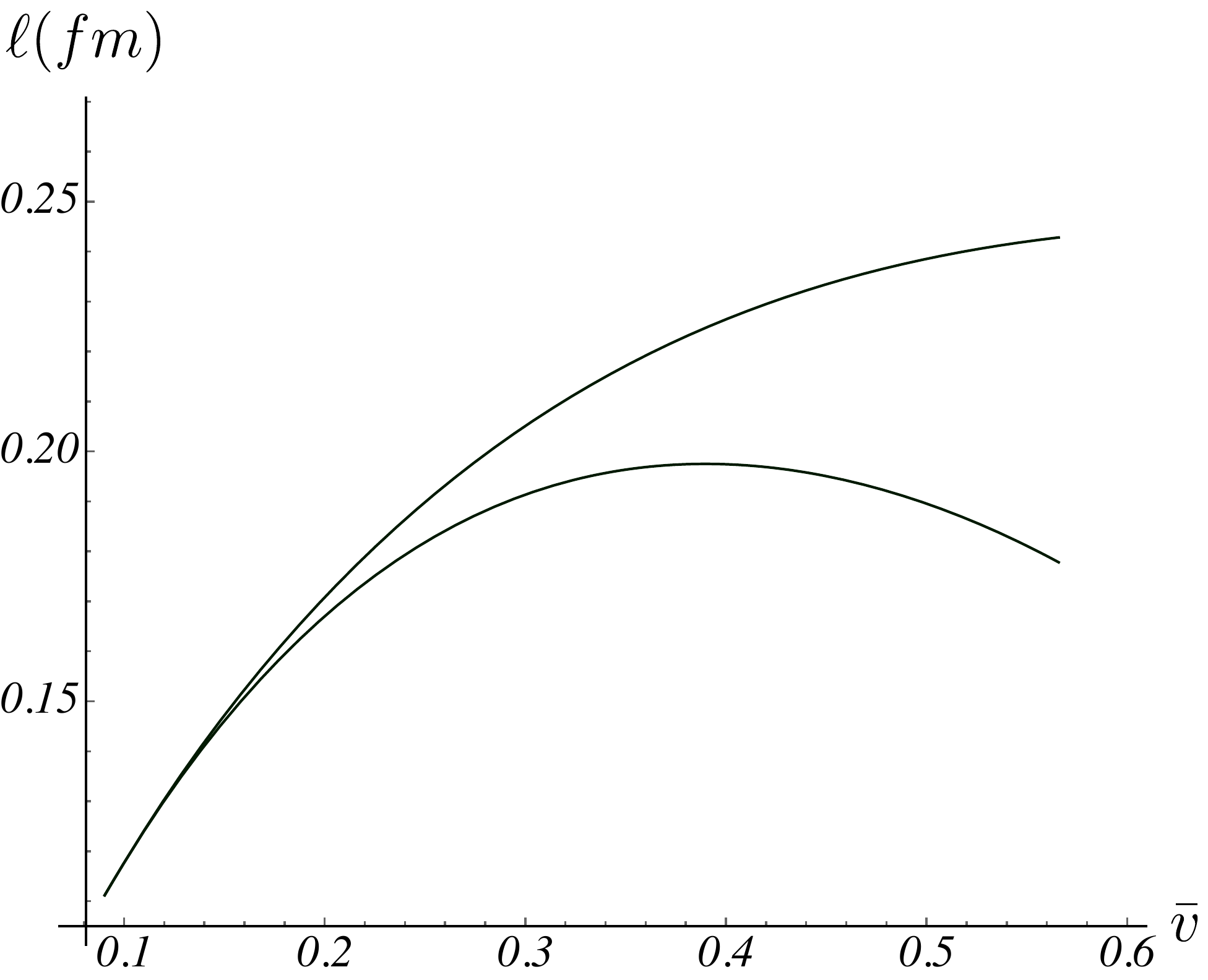}
\hspace{3cm}
\includegraphics[width=6.25cm]{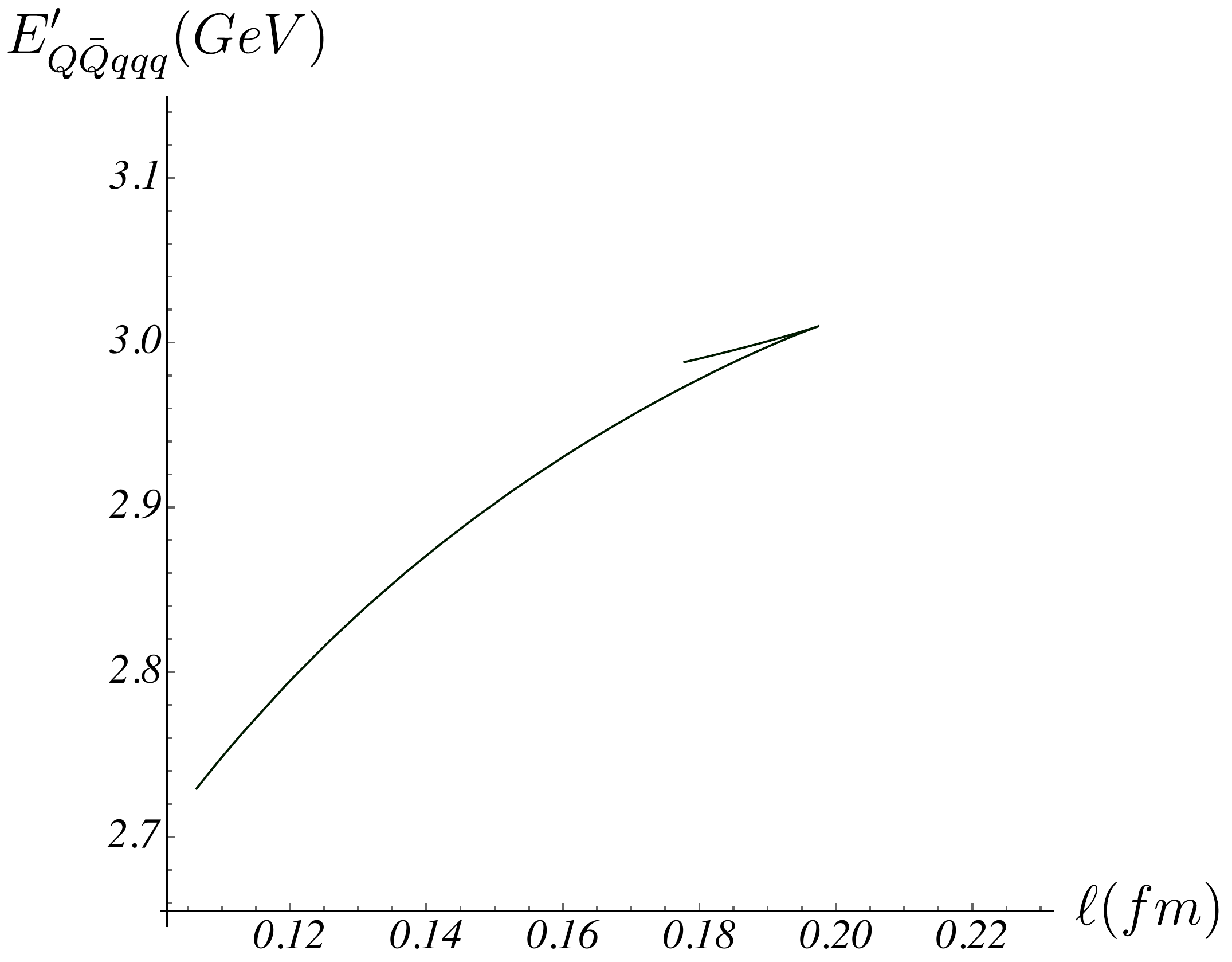}
\caption{{\small Left: $\ell$ as a function of $\bar v$. The lower and upper curves are associated with the configurations (c') and (c), respectively. Right: $E'_{\QQbqqq}$ vs $\ell$.}}
\label{E'c}
\end{figure}

A numerical calculation shows that, for $\k=\kv$, the function $\ell(\bar v)$ is not monotonically increasing on the interval $[\vs,\qs]$. Instead, it develops a local maximum close to $\bar v=0.4$. This means that such a configuration exists only if
    the distance $\ell$ does not exceed a critical value.\footnote{It is interesting to note that a similar situation arises when calculating the simplest connected string configuration in the AdS-like models at finite temperature \cite{djg,az3}.} Figure \ref{E'c} illustrates the distinct behaviors of $\ell(\bar v)$ for the two  
 configurations of Figure \ref{c52}. Here, we also include the result for $E'_{\QQbqqq}$ for the sake of completeness.


\end{document}